%% file: main.tex
\documentclass[10pt]{llncs}
\input{preamble}

\input{macros}

\title{HyperQB 2.0: A Bounded Model Checker for Hyperproperties}
\author{Tzu-Han Hsu\inst{1} 
\and Milad Rabizadeh\inst{1} 
\and Kenneth Rogale\inst{1} 
\and Fedor Filippov\inst{1} 
\and Marco A. de Oliveira Batista\inst{1} 
\and Cesar Sanchez\inst{2} 
\and Borzoo Bonakdarpour\inst{1}}

\institute{Michigan State University, USA\and IMDEA Software Institute, Spain}
\begin{document}
\maketitle

\input{abstract}

\input{intro}
\input{inputoutput}

\input{algo}
\input{experiments}

\input{related}
\input{concl}

%\section{Acknowledgment} 
%We would like to thank Tess Murphy and Lilly Yanke for their contributions to developing the \code{NuSMV} parser and QBF formula generator. 

\bibliographystyle{plain}
\bibliography{bibliography}

\newpage
\appendix
\input{appendix}

\end{document}

%% file: preamble.tex
\mainmatter
\usepackage{tcolorbox}
\tcbuselibrary{skins}

\tcbset{
  myproblem/.style={
    colback=gray!5,    % light gray background
    colframe=black!40, % gray border
    coltitle=black,
    fonttitle=\bfseries,
    boxrule=0.8pt,
    arc=2pt,
    outer arc=2pt,
    left=6pt,
    right=6pt,
    top=6pt,
    bottom=6pt,
    enhanced,
    sharp corners,
    attach boxed title to top left={xshift=0.5em,yshift=-2mm},
    boxed title style={colback=white, colframe=black!40, boxrule=0.5pt},
  }
}

\usepackage{listings} % in preamble
\tcbuselibrary{listings}
\newtcblisting{bnfbox}{
  colback=white,
  colframe=black!30,
  boxrule=0.5pt,
  listing only,
  listing options={
    basicstyle=\ttfamily\small,
    columns=fullflexible,
    keepspaces=true,
    frame=none,               % <-- remove inner box
    showstringspaces=false,
    numbers=none              % <-- turn off line numbers
  },
  left=6pt,right=6pt,top=0pt,bottom=0pt,arc=2pt
}

\lstdefinelanguage{NuSMV}{
  morekeywords={MODULE,VAR,ASSIGN,init,next,case,esac,TRUE,FALSE},
  sensitive=true,
  morecomment=[l]{--},
}

\usepackage[font=small,labelfont=bf]{caption}
\usepackage{tabulary}
\usepackage{fancyvrb}
\usepackage{multirow}
\usepackage{array}
\usepackage{booktabs}   %% For formal tables:
\usepackage{caption}
\usepackage{subcaption} %% For complex figures with subfigures/subcaptions
\usepackage{amsmath,amssymb}
\usepackage{amssymb}% http://ctan.org/pkg/amssymb
\usepackage{pifont}%
\usepackage[new]{old-arrows}
\usepackage{amsfonts}
\usepackage{xspace}
\usepackage[utf8]{inputenc}
\usepackage[T1]{fontenc}
\usepackage{array}
\usepackage{tabularx}
\usepackage{makecell}

\usepackage{hhline}
\usepackage{enumerate}
\usepackage{xcolor,colortbl}
\usepackage{paralist}
\usepackage{wrapfig}
\usepackage{listings}
\usepackage{tikz}

\usepackage[vlined,linesnumbered,ruled,resetcount]{algorithm2e}
\SetKwBlock{Repeat}{repeat}{}
\SetKwInOut{Initialization}{Initialization}

\usepackage{wrapfig}

\usepackage{todonotes}
\usepackage{ltl}

\usepackage{enumitem}
\usepackage{multicol}

\newcolumntype{M}[1]{>{\centering\arraybackslash}m{#1}}

\usepackage{diagbox}

\usepackage{stackrel}
\tikzset{
	>=stealth,
	possible world/.style={circle,draw,thick,align=center},
	real world/.style={double,circle,draw,thick,align=center},
}
\usetikzlibrary{arrows.meta, positioning}
\tikzset{%
	block/.style = {draw=\boxcolor,rectangle,thick,
		minimum height=2em, text width=10em, align=center},
	line/.style = {draw=\boxcolor, line width=4pt, 
		-{Triangle[length=10pt, width=10pt]}, shorten >=2pt, shorten <=2pt},
}

\usetikzlibrary{positioning,mindmap,automata,fit,backgrounds,shapes, 
arrows,shapes.misc,patterns}
\usetikzlibrary{decorations.pathmorphing}
\usetikzlibrary{decorations.markings}
\usetikzlibrary{calc}

\usepackage{graphicx}
\newsavebox{\nusmvbox}
\newsavebox{\grammarbox}
\newsavebox{\CLIbox}

\usepackage{tikz}
\usetikzlibrary{arrows, arrows.meta}
\usepackage[ruled,vlined,linesnumbered]{algorithm2e}
\usepackage{adjustbox}
\tikzset{arw/.style={>={Triangle[length=3mm,width=3mm]},line width=2mm,draw=gray}}

\tikzstyle{box} = [rectangle, minimum width=3cm, minimum height=1cm,text centered, text=white, draw=black, fill=black!50]
\usepackage{cite}
\usepackage{stmaryrd} %\llbracket for \sem

\usepackage{hyperref}
 \hypersetup{%
   colorlinks=true,           % use colored links
   allcolors=blue!70!black,   % use dark blue for all links
   pdfstartview=Fit,          % open PDF viewer with side-pane hidden.
   breaklinks=true,
   pdfauthor={T.-H. Hsu, C. Sánchez, B. Bonakdarpour},
   pdftitle={Bounded Checking for Hyperproperties}}
  
    \tikzset{every picture/.style=semithick,
 }%every axis plot/.append style=thick}
 \usepackage{cleveref}

%% file: macros.tex
\newcommand{\ourtool}{\code{HyperQB}$^{\code{2.0}}$\xspace}

\newcommand{\PC}{\textit{PC}}

\newcommand{\easim}{\textsf{SIM}_{\exists\forall}}
\newcommand{\aesim}{\textsf{SIM}_{\forall\exists}}

\newcommand{\GNI}{\textsf{\small GNI}\xspace}

\newcommand{\toolsmt}{\code{Z3}\xspace}
\newcommand{\toolquabs}{\code{QuAbS}\xspace}
\newcommand{\Yosys}{\code{Yosys}\xspace}
\newcommand{\mermaid}{\code{Mermaid}\xspace}

\newcommand{\ignore}[1]{}

\newcommand{\HyperLTL}{\text{HyperLTL}\xspace}
\newcommand{\AHLTL}{\text{A-HLTL}\xspace}

\newcommand{\MCHyper}{\textsf{\small MCHyper}\xspace}
\newcommand{\ABC}{\textsf{\small ABC}\xspace}

\newcommand{\quabs}{\textsf{\small QuAbS}\xspace}

\newcommand{\trace}{t}

\newcommand{\quant}{\mathbb{Q}}

\newcommand\donotshow[1]{}

\newcommand{\code}[1]{\textsf{\small #1}}
\newcommand{\typingfont}[1]{\texttt{#1}}

\newcommand{\qbf}[1]{\llbracket #1 \rrbracket}

\newcommand{\Tr}{\mathcal{T}}

\newcolumntype{K}[1]{>{\centering\arraybackslash}p{#1}}

\newcommand{\secret}{\mathit{high}}
\newcommand{\public}{\mathit{low}}

\pgfdeclarelayer{background}
\pgfdeclarelayer{foreground}
\pgfsetlayers{background,main,foreground}

\tikzset{
  invisible/.style={opacity=0, text opacity=0},
  visible on/.style={alt={#1{}{invisible}}},
  alt/.code args={<#1>#2#3}{%
    \alt<#1>{\pgfkeysalso{#2}}{\pgfkeysalso{#3}} % \pgfkeysalso doesn't change 
the path
  },
}

\tikzstyle{place}=[circle,thick,draw=blue!75,fill=blue!20,minimum size=6mm]
\tikzstyle{red place}=[place,draw=red!75,fill=red!20]
\tikzstyle{transition}=[rectangle,thick,draw=black, fill=black, 
minimum size=1mm]

\definecolor{mGreen}{rgb}{0,0.6,0}
\definecolor{mGray}{rgb}{0.5,0.5,0.5}
\definecolor{mPurple}{rgb}{0.58,0,0.82}
\definecolor{backgroundColour}{rgb}{0.95,0.95,0.92}

\lstdefinestyle{CStyle}{
    backgroundcolor=\color{backgroundColour},   
    commentstyle=\color{mGreen},
    keywordstyle=\color{magenta},
    numberstyle=\tiny\color{mGray},
    stringstyle=\color{mPurple},
    basicstyle=\footnotesize,
    breakatwhitespace=false,         
    breaklines=true,                 
    captionpos=b,                    
    keepspaces=true,                 
    numbers=left,                    
    numbersep=2pt,                  
    showspaces=false,                
    showstringspaces=false,
    showtabs=false,                  
    tabsize=2,
    language=C
}

\newcommand{\tupleof}[1]{\langle#1\rangle}

\newcommand{\OPT}{\mathit{opt}}
\newcommand{\PES}{\mathit{pes}}
\newcommand{\HOPT}{\mathit{hopt}}
\newcommand{\HPES}{\mathit{hpes}}

\newcommand{\Pmodels}{\models^{\PES}}
\newcommand{\Omodels}{\models^{\OPT}}
\newcommand{\HOmodels}{\models^{\HOPT}}
\newcommand{\HPmodels}{\models^{\HPES}}

\newcommand{\cmark}{\text{\ding{51}}}
\newcommand{\xmark}{\text{\ding{55}}}

\newcommand{\programgraph}{\mathcal{P}}
\newcommand{\locations}{\textit{Loc}}
\newcommand{\location}{\ell}

\newcommand{\effects}{\textit{Effect}\xspace}
\newcommand{\conditions}{\textit{Cond}\xspace}
\newcommand{\condition}{\textit{g}}
\newcommand{\transitions}{\longrightarrow}
\renewcommand{\trace}{t}

\newcommand{\HyperQube}{\code{HyperQB}\xspace}
\newcommand{\HyperQB}{\HyperQube}
\newcommand{\AutoHyper}{\code{AutoHyper}\xspace}

\newcommand{\Rust}{\code{Rust}\xspace}

\newcommand{\rh}{\textit{\rh}}
\newcommand{\RH}{\textit{\RH}}
\newcommand{\lh}{\textit{\lh}}
\newcommand{\LH}{\textit{\LH}}

\newcommand{\QBFall}[4]{\qbf{#4}^{#1}_{#2,#3}}
\newcommand{\QBFany}[3]{\QBFall{*}{#1}{#2}{#3}}

\newcommand{\boxcolor}{orange}

\newcommand{\nusmv}{\code{NuSMV}\xspace}
\newcommand{\verilog}{\code{Verilog}\xspace}

\newcommand{\better}[1]{{\cellcolor{yellow!60}\bf{#1}}}
\newcommand{\bettercex}[1]{{\cellcolor{green!60}\bf{#1}}}
\newcommand{\ocamlnum}[1]{{\cellcolor{gray!20}#1}}
\newcommand{\rustnum}[1]{{\cellcolor{gray!10}#1}}
\newcommand{\doublecheck}[1]{\textcolor{black}{#1}}
\newcommand{\forq}[1]{\underline{#1}}

\newcommand{\pgs}{\mathbb{P}}

%% file: abstract.tex
\vspace{-2mm}

\begin{abstract}
We introduce the tool \ourtool, the first highly efficient push-button bounded model checker (BMC) for hyperproperties.
% 
%Hyperproperties are system-wide properties that relate multiple
%computation traces, including many important information-flow
%security policies and consistency conditions.
% 
\HyperQB takes as input a model in \nusmv or \verilog and a formula expressed in the temporal logics \HyperLTL or \AHLTL.
The core decision procedures to implement BMC are SMT and QBF solvers, enabling verification of finite- and infinite-state programs.
\HyperQB offers command-line and standalone graphical, and web-based interfaces.
Based on the selection of either bug-hunting or synthesis, instances of counterexamples or path witnesses are returned.
The tool is entirely implemented in \Rust and we report on successful and effective model checking results for a rich set of experiments on a variety of case studies with rigorous performance comparison and contrast with similar tools. 

%\\[1em]
%\textbf{Data Availability Statement}: An artifact will be submitted to the AEC under EasyChair id 136.
%

\end{abstract}

%% file: intro.tex
\vspace{-8mm}
\section{Introduction}
\label{sec:intro}

Hyperproperties~\cite{cs10} are {\em system-wide} properties 
(rather than the property of individual execution traces).
They can express important information-flow security policies (e.g., non-interference~\cite{gm82}), consistency models in concurrent computing~\cite{bss18} (e.g., linearizability~\cite{hw90}), robustness conditions in cyber-physical 
systems~\cite{wzbp19}, and path planning in multi-agent systems.
The temporal logics HyperLTL~\cite{cfkmrs14} and A-HLTL~\cite{bcbfs21} extend LTL with explicit and simultaneous quantification over execution traces, describing relational properties of multiple synchronous and asynchronous traces, respectively.
For example, generalized noninterference (GNI) can be expressed by the following HyperLTL formula:
$$
\varphi_{\GNI} = \forall \pi_A. \forall \pi_B. \exists \pi_C. 
~X (\secret_{\pi_A} \leftrightarrow   \secret_{\pi_C})\
\land\ 	 
G\ (\public_{\pi_B} \leftrightarrow \public_{\pi_C}), 
$$
stipulates that for all traces $\pi_A$ and $\pi_B$, there must exists a 
$\pi_C$, 
such that
$\pi_{C}$ agrees on $\secret$ (i.e., {\em high-security} secret) with 
$\pi_{A}$,
and agrees on $\public$ (i.e., {\em low-security} observation) with 
$\pi_{B}$.
% their {\em high-security} value $\secret$ is different on $\pi_A$ and 
%$\pi_C$, but the {\em low-security} observation 
%$\public$ always stays the same on $\pi_B$ and $\pi_C$. 
%
Satisfying $\varphi_{\GNI}$ implies that an attacker cannot 
infer the high-security value by speculating the observable parts of a 
program.
%guess the secret value by observing the public information.

\begin{tcolorbox}[myproblem,title={Main Challenge}]
We currently lack serious tool support for verification of hyperproperties. Our objective is to develop an expressive, robust, user friendly, downloadable, efficient, and accessible model checking tool accompanied by a rich set of benchmarks and case studies. 
\end{tcolorbox}

\begin{figure}[t]
	\centering
		\vspace{-2mm}
    \includegraphics[width=1\linewidth] {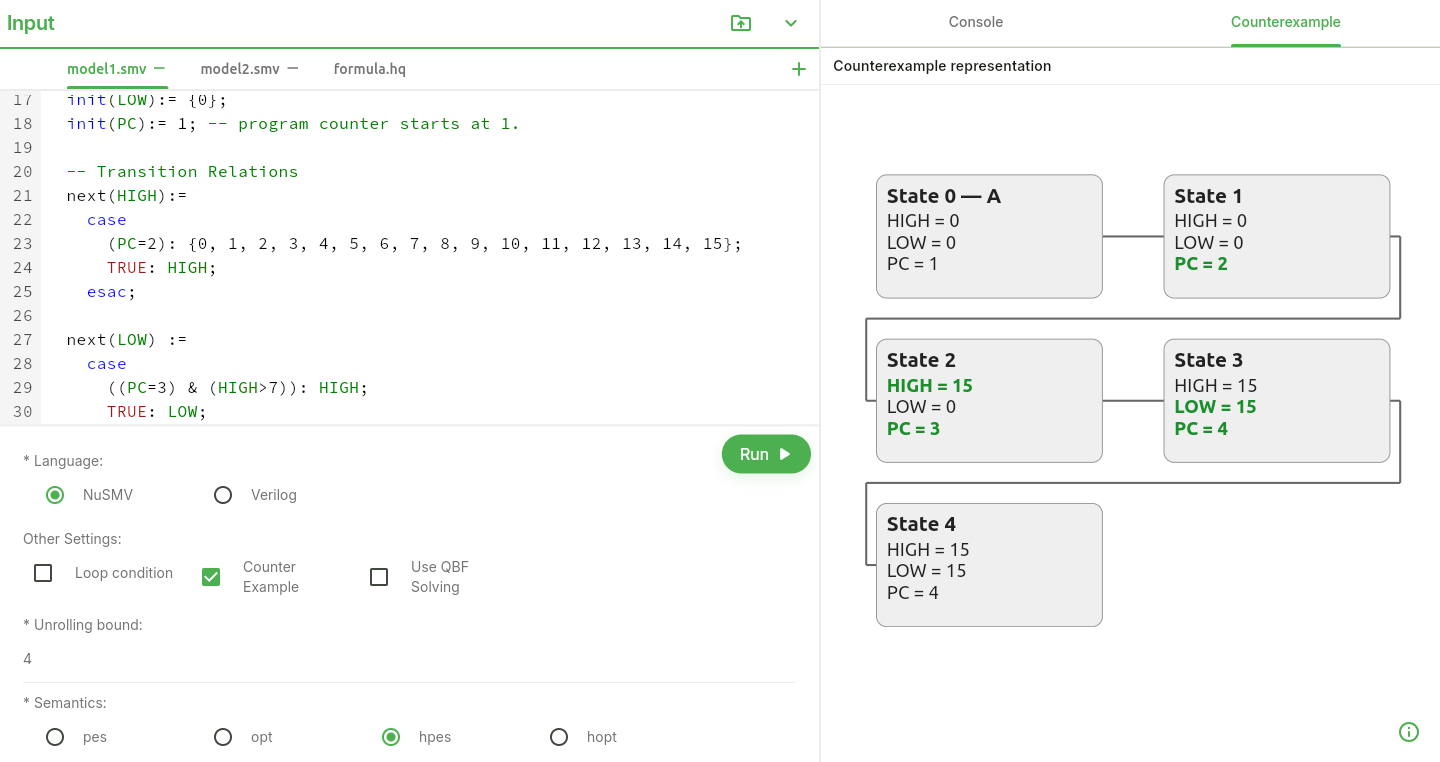}
	\caption{The standalone and web-based GUI for \ourtool demonstrating counterexample generation on a \nusmv model.
	}
	\label{fig:placeholder}
	\vspace{-5mm}
\end{figure}

This paper introduces \ourtool, a {\em bounded model checker} (BMC) for hyperproperties based on the QBF-based techniques introduced in~\cite{hsb21,hbfs23,hssb23}\footnote{The tool and documentation are available at \url{https://cse.msu.edu/tart/tools}.}.

\vspace{-3mm}
\paragraph{Key features.} In a nutshell, \ourtool has the following features:

\begin{itemize}
	
	\item It provides three user interfaces: (1) command line, (2) desktop standalone GUI, and (3) web-based playground for remote access (see~\Cref{fig:placeholder}).

	\item The inputs to \HyperQB are (1) a set of models (up to one per trace quantifier) in either \nusmv~\cite{cimatti1999nusmv} or \verilog~\cite{2005Verilog}, (2) a specification in HyperLTL or A-HLTL, (3), the choice of bounded semantics, and (4) the unrolling bound $k \geq 1$.
	
	\item The possible outputs of \HyperQB are: \texttt{sat}, \texttt{unsat}, or \texttt{unknown}. If the model does not satisfy the formula, a counterexample (either textual for command-line or visualized in the GUI) is generated (see~\Cref{fig:placeholder}).
\end{itemize}

\begin{figure}[htb!]
	\centering

	\includegraphics[width=\textwidth]{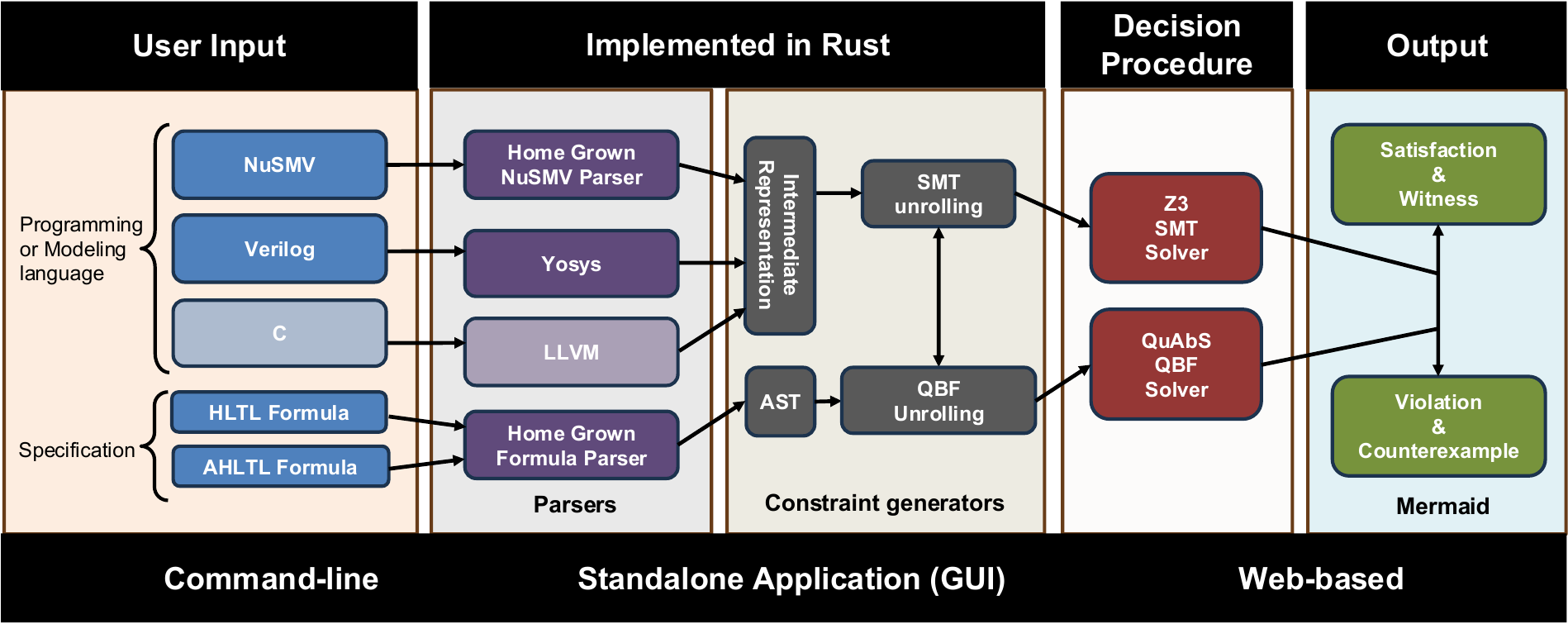}
		\vspace{-6mm}
			\caption{Overall architecture of \ourtool.}
	\label{fig:flowchart}
	\vspace{-7mm}
\end{figure}

\vspace{-5mm}
\paragraph{Overall architecture.}
Figure~\ref{fig:flowchart} shows the overall architecture of \ourtool implemented in \Rust.
We developed parsers for \HyperLTL, \AHLTL, and \nusmv.
\verilog programs are parsed using the tool \Yosys~\cite{YosysHQ}.
The \code{C} $\rightarrow$ \code{LLVM} path is currently under development.
These parsers translate \nusmv and \verilog models to a unified abstract intermediate representation as a {\em program graph}~\cite{bk08-book} (i.e., a transition relation with data).
Next, the program graph of models and the abstract syntax tree of the specification are handed to another home-grown
component for BMC unrolling and constraint generation.
\ourtool supports both SMT and QBF constraint generation, enabling verification of finite- and infinite-state programs.
Finally, we run the SMT-solver \toolsmt~\cite{dmb08} or the QBF-solver  \quabs~\cite{t19} to check the satisfiability of the unrolled 
instance and interpret the output.
\ourtool also incorporates the loop conditions identified in~\cite{hssb23} to ensure completeness of model checking for certain fragments of \HyperLTL.
We note that besides verification, \HyperQube can also be used for synthesis through returning witnesses to existential quantifiers in the input \HyperLTL and \AHLTL formulas.

\begin{tcolorbox}[myproblem,title={Comparison to \HyperQB 1.0}]
% [title=Comparison to \HyperQB 1.0]
An earlier prototype version of \HyperQB was developed to support the experimental evaluation in~\cite{hsb21}. {\bf No tool paper has ever been published for HyperQB}.
Developing \ourtool required massive theoretical and engineering effort and has the following unique and new features:
  
\begin{itemize}
	
	\item End-to-end implementation in {\bf Rust}.
	
	\item {\bf SMT-based} decision procedure (in addition to the original QBF-based), enabling verification of {\em infinite-state} programs.
	
	\item Implementation of {\bf loop conditions} to support BMC completeness.
	
	\item {\bf Verilog} as an additional language for hardware security applications.
 
	\item {\bf A-HLTL} specifications to support asynchronous hyperproperties.
	
	\item Significant expansion of {\bf benchmarks}, creating a wealth of examples for the research community (those in~\Cref{tab:result-ahltl,tab:results_verilog,tab:performance,tab:result-hltl2} are all new).
	
	\item Significant improvement in {\bf performance}.
	
	\item Graphical standalone and web-based {\bf user interfaces}.
	
\end{itemize}
\end{tcolorbox}

We also present a rigorous experimental evaluation and compare \ourtool to our preliminary findings in~\cite{hsb21,hbfs23,hssb23} as well as the tool \AutoHyper~\cite{bf23}.
Our experimental evaluation includes a wide range of case studies,
such as information-flow security, linearizability in concurrent
data structures, path planning for robots, co-termination, 
deniability, intransitivity of non-interference, and \linebreak secrecy-preserving mapping synthesis.
Our evaluation shows that our technique is effective and efficient
in identifying bugs in several prominent examples.

\vspace{-2mm}
\paragraph{Organization.} The rest of the paper is organized as follows.
\Cref{sec:overview} discusses the user interface and input/output features.
The core algorithmic backbone and implementation are described in~\Cref{sec:algorithm}.
\Cref{sec:eval} presents detailed empirical evaluation. 
Related work is discussed in~\Cref{sec:related}.
Finally, we make concluding remarks and discuss future work in~\Cref{sec:concl}. 

%% file: inputoutput.tex
\section{Input/Output Features of HyperQB}
\label{sec:overview}

\begin{figure}[t]
	\centering
	\input{figs/nusmv}
	\scalebox{0.83}{\usebox{\nusmvbox}} 
	\caption{A NuSMV model with implicit leak from \typingfont{high} to \typingfont{low}.}
	\vspace{-3mm}
	\label{fig:nusmv-model}
\end{figure}

%This section describes the input and output interfaces of \HyperQB, detailing how users provide models and specifications, configure parameters, and interpret the tool’s results.

\vspace{-7mm}
\subsection{Modeling Languages}\label{subsec:modeling}

\begin{wrapfigure}{r}{0.44 \textwidth}
	\centering
	\vspace{-22mm}
	\scalebox{0.68}{
		\input{figs/KS}
	}
	\caption{Transition relation of~\Cref{fig:nusmv-model}.}
	\vspace{-5mm}
	\label{fig:nusmv-ts}
\end{wrapfigure}
\HyperQB accepts input models in \nusmv~\cite{cimatti1999nusmv} and the HDL \verilog~\cite{2005Verilog}.
For \nusmv, we have built a parser to extract transition relations.
% \tzuhan{details in Section 3.2 about how this TS is built.}
%
For example, the \nusmv model in~\Cref{fig:nusmv-model} encodes a system where the low-security variable \textit{low} is updated under conditions depending on the high-security variable \textit{high}. 
Although no explicit flow from \textit{high} to \textit{low} exists, the model induces an implicit leak and violates formula $\varphi_{\mathit{\GNI}}$ introduced in~\Cref{sec:intro}.
The corresponding transition relation in~\Cref{fig:nusmv-ts} shows two possible executions, branching on the value of \textit{high}, showing that an attacker observing \textit{low} as true at $\PC=3$ can immediately infer that the secret input \textit{high} must be true.

For \verilog, the top-level module must be provided, where the open-source tool \Yosys~\cite{YosysHQ} translate the HDL into its SMT-LIB representation.
Through \Yosys, we inherit support for a substantial fragment of the 2005 \verilog standard~\cite{2005Verilog}.

\vspace{-4mm}
\subsection{Input Specification and Semantics}
\label{sec:spec_semantics}
\vspace{-2mm}

\paragraph{Specification language.} \HyperQB supports input specification in two temporal logics: (1)  \HyperLTL~\cite{cfkmrs14} for {\em synchronous}, and (2) \AHLTL~\cite{bcbfs21} for {\em asynchronous} hyperproperties.
To motivate these two variations, consider~\Cref{fig:nusmv-ts} and traces $t_1 = s_0s_1s_2$ and $t_2 = s_0s_3s_4$.
One way to evaluate the security policy is to pair $s_1$ with $s_3$ and then $s_2$ with $s_4$ (i.e., lock-step synchronous evaluation of traces $t_1$ an $t_2$). This is how \HyperLTL would evaluate $t_1$ and $t_2$.
Alternatively, one may need to evaluate $t_1$ and $t_2$, where $t_1$ stutters at $s_1$ (illustrated here by the {red dashed} alignment).
This is how \AHLTL evaluates $t_1$ and $t_2$ asynchronously.
More specifically, traces $t_1$ and $t_2$ satisfy formula $\varphi_{eq}= \mathbf{G}(\PC_{\pi_A} = \PC_{\pi_B})$ in HyperLTL but violate the same formula in \AHLTL for the red alignment.

\begin{figure}[t]
	\centering
	\begin{lrbox}{\grammarbox}
		\input{figs/grammar}
	\end{lrbox}
	\scalebox{0.75}{\usebox{\grammarbox}} 
	\caption{Grammar of \HyperLTL/\AHLTL input formulas (full grammar in~\Cref{appendix:fullgrammar}).}
	\label{fig:grammar}
	\vspace{-3mm}
\end{figure}

Grammars of these two logics within \HyperQB are shown in~\Cref{fig:grammar}. 
The grammar for each logic has two rules \typingfont{path\_formula}, which quantifies over traces, and \typingfont{traj\_formula}, which quantifies over trajectories (asynchronous alignments). 
When the input logic is \typingfont{AHLTL\_rec}, the grammar temporal core is \typingfont{inner\_AHLTL} (otherwise,  \typingfont{inner\_HLTL}) followed by rules \typingfont{hform} and \typingfont{aform} for paths and trajectories. Expressions are inductively generated by Boolean connectives, arithmetic comparisons, and temporal operations (see \typingfont{binary\_op} and \typingfont{unary\_op}). 
%
%This hierarchy enforces precise binding and unambiguous parsing, providing a rigorous grammar for the input formula.

% also temporal operators such as \emph{until} (\typingfont{U}) and \emph{release} (\typingfont{R}). 
%
% Formulae are further refined through unary operators such as \emph{globally} (\typingfont{G}), \emph{eventually} (\typingfont{F}), \emph{next} (\typingfont{X}). 
% parenthesization, and atomic propositions, where atoms consist of indexed identifiers, constants, or numbers. 
%
% The entry rule requires a specification to begin with a path formula and extend to the end, yielding a rigorous grammar as part of the user inputs of \ourtool.
% \tzuhan{Is~\Cref{fig:grammar} a good way to present grammar?}

\begin{table}[b]
\vspace{-7mm}
	\centering
    \caption{Bounded semantics for a HyperLTL/A-HLTL formula $\varphi$.}
	\begin{tabular}{m{1cm} m{4.55cm} @{\hspace{1.7em}} m{5.6cm}}
		\toprule
		\textbf{Sem} & \textbf{Intuition} & \textbf{Infinite Inference} \\
		\midrule
		% \textit{Pes}
		\texttt{-pes}
		& $\varphi$ is declared \emph{false} unless it is witnessed as true within \texttt{k} steps.  & If $(\Tr,\Pi,0)\Pmodels_k\varphi$, then $(\Tr,\Pi,0)\models\varphi$ \\
		\addlinespace
		% \textit{Opt} 
		\texttt{-opt}
		& $\varphi$ is declared \emph{true} unless it is witnessed as false within \texttt{k} steps.  
		& If $(\Tr,\Pi,0)\not\Omodels_k\varphi$,  then $(\Tr,\Pi,0)\not\models\varphi$ \\
		\addlinespace
		% \makecell[l]{\textit{Halting}\\\textit{Pessimistic}} 
		% \textit{H-Opt}
		\texttt{-hopt}
		& $\varphi$ is declared \emph{false} unless it is witnessed as true before \texttt{halt}. & If $(\Tr,\Pi,0)\HPmodels_k\varphi$, then $(\Tr,\Pi,0)\models\varphi$ \\
		\addlinespace
		% \makecell[l]{\textit{Halting}\\\textit{Optimistic}} 
		% \textit{H-Pes}
		\texttt{-hpes}
		& $\varphi$ is declared \emph{true} unless it is witnessed as false before \texttt{halt}. & If $(\Tr,\Pi,0)\not\HOmodels_k\varphi$, then $(\Tr,\Pi,0)\not\models\varphi$ \\
		\bottomrule
	\end{tabular}
	\vspace{-8mm}
	\label{tab:semantics}
\end{table}

\vspace{-3mm}
\paragraph{Bounded semantics.} \HyperQB incorporates all the bounded semantics introduced in~\cite{hsb21} (see in~\Cref{tab:semantics}).%, providing a diverse set of approaches to reason about \HyperLTL and \AHLTL formulas over bounded traces.
These bounded variations guarantee sound inference with respect to the unbounded semantics, ensuring that no incorrect conclusions are drawn from finite exploration. 
This enables verification within a fixed bound $k \geq 0$ or until program termination (denoted as \typingfont{halt}). 
For instance, to verify whether the model in~\Cref{fig:nusmv-ts} satisfies $\varphi_{\GNI}$, we first negate the property and apply the pessimistic semantics (\typingfont{-pes}) for bug hunting.
In this setting, $\varphi_{\GNI}$ is reported as violated because a counterexample is found within the unrolling bound $k$ (i.e., aligned with the first case in~\Cref{tab:semantics}).
%
% The choice of semantics depends on the verification objective: pessimistic (\typingfont{-pes}) for bug-hunting (detecting violations), optimistic (\typingfont{-opt}) for witness finding (establishing that a property holds along a finite prefix), halting-pessimistic (\typingfont{-hpes}), or halting-optimistic (\typingfont{-hopt}) semantics for terminating programs (ensuring conclusions remain valid even under larger bounds).
%
%We will elaborate on these choices when discussing the case studies in~\Cref{sec:eval}.\borzoo{I don't think we do this.}

%Although the transition system in~\Cref{fig:nusmv-ts} is acyclic (since the NuSMV model has no loops), \HyperLTL can also be evaluated over programs with loops. 
%%
%In such cases, executions may be infinite, but bounded model checking leverages loop conditions to reason soundly about their behavior. 
%%
%Thus, even for infinite traces, one can conclude the satisfaction of global properties such as $\varphi_{eq}$ under the established \emph{loop conditions}~\cite{hsu2023efficient}.

\begin{wraptable}{r}{0.48\linewidth}
	\centering
    \caption{Categories of loop conditions.}
	\begin{tabular}{p{1.7cm} p{3.3cm}}
		\toprule
		\textbf{Case} & \textbf{Infinite Inference} \\
		\midrule
		$\forall_{\text{small}} \; \exists_{\text{big}}$ 
		& $\aesim \to\;\; \models \forall \exists~ \mathbf{G} \psi$ \\
		
		$\forall_{\text{big}} \; \exists_{\text{small}}$ 
		& $\aesim \to\;\; \models \forall \exists~ \mathbf{G} \psi$ \\
		
		$\exists_{\text{small}} \; \forall_{\text{big}}$ 
		& $\easim \to\;\; \models \exists \forall~ \mathbf{G} \psi$ \\
		
		$\exists_{\text{big}} \; \forall_{\text{small}}$ 
		& $\easim \to\;\; \models \exists \forall~ \mathbf{G} \psi$ \\
		\bottomrule
	\end{tabular}
	\vspace{-7mm}
	\label{tab:loopcond}
\end{wraptable}
\paragraph{Loop conditions.}
Similar to the classic BMC approach~\cite{cbrz01}, \ourtool implements loop conditions to ensure completeness.
\Cref{tab:loopcond} shows the loop conditions of fragments of  \HyperLTL and the recommended relative size of models for better performance. 
The core idea is to employ a simulation encoding in which finite unrolling suffices to conclude the satisfaction of a \emph{globally} temporal formula in infinity~\cite{hssb23}. 
For example in~\Cref{subsec:modeling}, if we extend the \nusmv model with an additional line, where in state $\PC=3$ the counter is reset to $\PC=1$ (line~\Cref{line:add-loop} in~\Cref{fig:nusmv-model}), then the extracted transition system will contain cycles (shown as the gray dashed transitions in~\Cref{fig:nusmv-ts}).
In this setting, verifying $\forall \pi_A.~ \exists \pi_B.~ \mathbf{G}(\PC_{\pi_A} = \PC_{\pi_B})$ 
requires checking the satisfiability of $\aesim$, where the corresponding models are of equal size. 
% requires establishing that the property holds over an infinite execution horizon.  
%
% That is, positive outcomes (e.g., $\models \forall\exists~ \typingfont{G} \varphi$) can be established

%
%By integrating this simulation-based encoding, we enable automatic exploitation of these fragments during verification, thereby allowing \HyperQB to handle a broad range of verification cases for programs with loops (see~\Cref{sec:eval}).

\vspace{-3mm}
\subsection{User Interface, Tool Usage, and Output}
\vspace{-1mm}

\paragraph{Tool interface and output.}
%\ourtool requires the user to specify the model, the formula, the unrolling bound, and the chosen semantics. For \verilog programs, the name of the top-level module must also be provided.
%
The output of \HyperQB consists of the verification result (i.e., either \texttt{sat}, \texttt{unsat}, or \texttt{unknown}) produced by the selected backend solver (i.e., \toolsmt~\cite{dmb08} or \toolquabs~\cite{t19}), accompanied by a counterexample when requested.
% The solver returns \texttt{sat}, \texttt{unsat}, or \texttt{unknown}.
The \texttt{unknown} outcome indicates that the solver was not able to decide the satisfiability of the instance.
When the counterexample option (\texttt{-c}) is enabled and the result is unsatisfiable, \HyperQB generates a counterexample textual trace.
%
%This counterexample constitutes a witness to the violation of the hyperproperty.

\begin{figure}[t]
\centering
    \begin{lrbox}{\CLIbox}
    \input{figs/CLI}
    \end{lrbox}
    \scalebox{0.75}{\usebox{\CLIbox}}
    \caption{\ourtool command-line interface usage and arguments.}
    \label{fig:hyperqb-cli}
    \vspace{-3mm}
\end{figure}

\vspace{-2mm}
\paragraph{Command-line usage example.}
An overview of \HyperQB’s CLI usage, arguments, and an example usage is given in~\Cref{fig:hyperqb-cli}.
In the example, the given linearizability formula (\typingfont{-f lin.hq}) is of the form $\forall\exists$, so one model per quantifier is required: \typingfont{-n snark\_conc.smv} \typingfont{snark\_seq.smv}.
%
% In this case two different models are provided, and 
The  unrolling bound is specified by a non-negative integer (\typingfont{-k 18}), and the selected bounded semantics is halting-pessimistic (\typingfont{-s hpes}). 
As the \texttt{-c} option is provided, \HyperQB will return the verdict (\typingfont{unsat} in this case), accompanied by a counterexample.

\vspace{-2mm}
\paragraph{Graphical User Interface (GUI).}
\HyperQB also offers a GUI (see~\Cref{fig:placeholder}), available as both {\em standalone desktop} and {\em web-based} (playground at \url{https://hyperqb.egr.msu.edu/}) applications. 
% is oriented towards interactive use.
% providing identical functionality as the CLI.
%
In the GUI, inputs are organized into tabs, enabling users to manage multiple model files and the formula within a single session. 
Different modeling languages can be selected (either \nusmv or \verilog), and configuration parameters are fully accessible (unrolling bound, semantics, etc). 
Upon execution, the solver output is shown in a dedicated pane.
When counterexample generation is enabled, the violating trace is shown as a graphical sequence of states, each annotated with the color-coded values of the relevant variables using the \mermaid~\cite{mermaid} diagramming tool format (see~\Cref{fig:placeholder}).
%
%The format emphasizes variables that change, providing a clear visualization of the system’s evolution and offering valuable support for debugging and explaining property violations (see~\Cref{fig:placeholder}).

%% file: figs/nusmv.tex
\begin{lrbox}{\nusmvbox}
\begin{minipage}[t]{0.45\textwidth}
\begin{lstlisting}[language=NuSMV, firstnumber=1]
MODULE main
VAR
  low:  boolean; high: boolean; 
  halt: boolean; PC:   1..3;
ASSIGN
  -- Initial Conditions
  init(low)  := FALSE;
  init(high) := FALSE;
  init(halt) := FALSE;
  init(PC)   := 1;

  -- Transition Relations
  next(high) :=
    case
      (PC=1): {TRUE, FALSE};
      TRUE: high;
    esac;

\end{lstlisting}
\end{minipage}
\hfill\hspace{7mm}
\begin{minipage}[t]{0.47\textwidth}
\begin{lstlisting}[language=NuSMV, firstnumber=19, escapeinside={(*@}{@*)}]
  next(low) :=
    case
      (PC=2 & high): TRUE;
      TRUE: low;
    esac;
  next(halt) :=
    case
      (PC=3): TRUE;
      TRUE: halt;
    esac;
    
  next(PC) :=
    case
      (PC<3): PC+1;
      -- (PC=3): 1; // if looping (*@\label{line:add-loop}@*)
      TRUE: PC;
    esac;
\end{lstlisting}
\end{minipage}
\end{lrbox}

%% file: figs/KS.tex
\begin{tikzpicture}[
    >=Latex,
    state/.style={
        % ellipse, draw=black!70, very thick,
        rectangle, rounded corners=5pt,
        draw=black, thin,
        % fill=black!6, 
        fill=blue!6,
        align=center,
        minimum width=20mm, minimum height=15mm,
        inner sep=2.5pt
    },
    lbl/.style={font=\scriptsize,inner sep=0pt},
    x=1mm, y=1mm
    ]
    
    % --- column x-positions (left and right branch) ---
    \def\xL{-15}   % left column x
    \def\xR{ 15}   % right column x
    \def\yTop{ 18} % s0 y
    \def\dy{ -22}  % vertical gap
    
    % s0 at top center
    \node[state] (s0) at (0,\yTop) {%
        \(\neg low,~\neg high,\)\\
        \(PC = 1,\)\\
        \(\neg halt\)};
    
    % left branch: s1 -> s2 (downwards)
    \node[state] (s1) at (\xL,\yTop+\dy) {%
        \(\neg low,~high,\)\\
        \(PC = 2,\)\\
        \(\neg halt\)};
    \node[state] (s2) at (\xL,\yTop+2*\dy) {%
        \(low,~high,\)\\
        \(PC=3,\)\\
        \(halt\)};
    
    % right branch: s3 -> s4 (downwards)
    \node[state] (s3) at (\xR,\yTop+\dy) {%
        \(\neg low,~\neg high,\)\\
        \(PC = 2,\)\\
        \(\neg halt\)};
    \node[state] (s4) at (\xR,\yTop+2*\dy) {%
        \(\neg low,~\neg high,\)\\
        \(PC = 3,\)\\
        \(halt\)};
    
    % labels
    \node[lbl] at ($(s0.north west)+(0,2mm)$) {\normalsize $s_0$};
    \node[lbl] at ($(s1.north west)+(0,2mm)$) {\normalsize $s_1$};
    \node[lbl] at ($(s2.north west)+(0,2mm)$) {\normalsize $s_2$};
    \node[lbl] at ($(s3.north east)+(0mm,1.5mm)$) {\normalsize $s_3$};
    \node[lbl] at ($(s4.north east)+(0mm,1.5mm)$) {\normalsize $s_4$};
    
    % initial arrow & title
    % \draw[->] ($(s0.west)+(-12mm,0)$) -- (s0.west);
    % \node[lbl] at ($(s0.west)+(-18mm,0)$) {$K_{\mathrm{exp}}:$};
    
    % transitions (downward)
    \draw[->, line width=0.9pt] (s0) -- (s1);
    \draw[->, line width=0.9pt] (s1) -- (s2);
    \draw[->, line width=0.9pt] (s0) -- (s3);
    \draw[->, line width=0.9pt] (s3) -- (s4);
    
    \draw[red, dashed, very thick] (s1) -- (s3);
    \draw[red, dashed, very thick] (s1) -- (s4);
    \draw[red, dashed, very thick] (s2) -- (s4);

    \draw[->, gray, dashed, line width=0.9pt] (s2.west) to[bend left=95] (s0.west);
    \draw[->, gray, dashed, line width=0.9pt] (s4.east) to[bend right=95] (s0.east);
    
\end{tikzpicture}

%% file: figs/grammar.tex
\begin{bnfbox}
path_formula ::=  ("forall" | "exists") pid . form_rec
traj_formula ::= ("A" | "E" ) tid . AHLTL_rec
    
form_rec     ::= path_formula | traj_formula | inner_HLTL
AHLTL_rec    ::= traj_formula | inner_AHLTL
    
inner_HLTL   ::= hform          inner_AHLTL  ::= aform
    
hform ::= hform binary_op hform | unary_op hform | h_atom 
aform ::= aform binary_op aform | unary_op aform | a_atom 
    
h_atom ::= id[pid] | constant | number | "(" hform ")"
a_atom ::= id[pid][tid] | constant | number | "(" aform ")"
    
binary_op ::= "U" | "R" | "=" | "->" | "&" | "|"
unary_op  ::= "G" | "F" | "X" | "~"
    
constant ::= "TRUE" | "FALSE"    number ::= [0-9]+ | "#b" (0|1)+
\end{bnfbox}

%% file: figs/CLI.tex
\begin{tcolorbox}[
    colback=white,
    colframe=black!30,
    fonttitle=\bfseries,
    boxrule=0.6pt,
    % sharp corners,arc=2pt
    % enhanced,
    ]
    \textbf{Synopsis} \\
    \texttt{hyperqb (-n|-v) <models> -f <formula> -k <int> -s <sem> [options]}
    
    \vspace{0.8em}
    \textbf{Arguments}
    
    \begin{tabularx}{\linewidth}{@{}lX@{}}
        
        \multicolumn{2}{l}{Required} \\[0.2em]
        \texttt{-n <files> \textbar\ -v <files>} & List of model files (NuSMV/Verilog). \\
        \texttt{-f <file>} & Formula file (\texttt{.hq}). \\
        \texttt{-t <name>} & Top module (required for Verilog only). \\ [0.4em]
        \texttt{-k <int> | -l} & Unrolling bound (steps) or loop condition. \\
        \texttt{-s <pes|opt|hpes|hopt>} & Choice of the bounded semantics. \\[0.6em]
        
        \multicolumn{2}{l}{Optional} \\[0.2em]
        \texttt{-m <int>} & Trajectory bound. \\
        \texttt{-c} & Emit counterexample for \textit{unsat} outcome. \\
        \texttt{-q} & Use QuAbS QBF solver (default: Z3). \\
    \end{tabularx}
    
    \vspace{0.8em}
    \textbf{Example Usage} \\
    \texttt{hyperqb -n snark\_conc.smv snark\_seq.smv -f lin.hq -k 18 -s hpes -c}
\end{tcolorbox}

%% file: algo.tex
\vspace{-3mm}
\section{Algorithmic Backbone and Implementation}
\label{sec:algorithm}
\vspace{-2mm}

% This section presents an overview of the internal technical implementation of \HyperQB to enable SMT/QBF-based BMC.
This section outlines the technical details of our \Rust implementation.
%.
The core is a streamlined intermediate representation (IR) that is both precise and flexible.
Both \nusmv and \verilog models are parsed into this IR, which is then compiled into solver-ready SMT/QBF encodings.
The same IR also supports the construction of simulation-based loop conditions.
{
\begin{tcolorbox}[myproblem,title={SMT vs. QBF Encodings}]

We only describe the SMT encoding in this section. Every SMT encoding can be systematically transformed into an equivalent QBF encoding via standard \emph{bit-blasting}, where each bounded integer variable $i$ is represented using $\lfloor \log(i) \rfloor + 1$ Boolean variables.
\end{tcolorbox}

\vspace{-3mm}
\subsection{Intermediate Representation (IR) of Input Models}
\label{sec:IR}

We use program graphs as the unified IR of input languages after parsing. 
A \emph{program graph} is a tuple $
\programgraph = \tupleof{\locations, \effects, \conditions, \transitions, \locations_0, \condition_0}$,
where $Loc$ is a finite set of control locations, $\effects$ maps a variable evaluation to an updated evaluation, $\conditions$ is a set of conditions over variables, $\longrightarrow \subseteq \locations \times \conditions \times \locations$ is the set of edges that represent control-flow transitions, $\locations_0 \subseteq \locations$ is a set of initial locations, and $\condition_0$ is the initial condition.  
A \emph{trace} $\trace$ of a program graph is a sequence obtained by following transitions from an initial location $\location_0 \in \locations_0$ under a variable valuation that satisfies the initial condition $\condition_0$. 
Concretely, a trace records the sequence of conditions encountered along a path, together with the corresponding variable valuations as updated by $\effects$.
%
%\paragraph{Effect Semantics.} 
%Unlike the definition in~\cite{baier2008principles}, where the \emph{effect} is modeled as a function over actions, 
%in the context of BMC unrolling we are only concerned with pre--post conditions. 
We define the effect function as a mapping over variables such that, 
for every assignment $\mathit{Effect}(x) = e$, the post-state variable $x'$ satisfies $x' = \llbracket e \rrbracket_{i}$, where $\llbracket e \rrbracket_i$ denotes the evaluation of expression $e$ under the variable valuation at step $i$.
%
% We use $\tracesof{\programgraph}$ to denote the set of all possible traces of $\programgraph$.
% Traces thus represent possible executions of the modeled program.

A program graph provides an abstract yet precise representation of program behavior: control locations correspond to program points, edges represent conditional checks, and the effect function specifies how variables evolve.
%
%\HyperQB adopts the program graph as its intermediate representation and encodes it directly in Z3, using the robust \Rust API to construct \emph{abstract syntax trees} (ASTs).
%
Models in \nusmv and \verilog are parsed and compiled into a program-graph IR, which is then materialized as a SMT-LIB–equivalent \emph{abstract syntax trees} (AST), enabling \emph{infinite state spaces} arising from unbounded integers, arrays, or data domains.
Note that it is unnecessary to explicitly track program locations in the AST, since pre- and post-conditions are defined over variable evaluations.
Consequently, the translation from \nusmv or \verilog to the IR omits explicit location information $\locations$.
% \tzuhan{how about we show the data structure of this SMVEnv? or at least how the variable types are defined?}
%
%This choice of IR offers both generality and expressiveness. Program graphs naturally capture the operational semantics of concurrent imperative programs and, unlike finite-state automata, readily accommodate \emph{infinite state spaces} arising from unbounded integers, arrays, or data domains. 
%
This represents a significant improvement over the prototype in~\cite{hsb21}, where the QBF-based IR required bit-blasted bounded integers.
%
%By grounding the IR in Z3, \HyperQB obtains a uniform, solver-ready representation that can be directly leveraged for verification. This design not only enables reasoning about a wide spectrum of systems through a single backend, but also ensures seamless interoperability with other modeling frameworks widely used in the model checking literature, such as Kripke structures and transition systems.

\vspace{-2mm}
\subsection{Translating NuSMV to IR} \label{subsec:NuSMv2IR}
%We next detail how \HyperQB processes its supported input languages, NuSMV and Verilog. 
%
%\HyperQB provides dedicated parsers for both \nusmv and \verilog, translating models into Z3 ASTs. 
%
%This produces an SMT-LIB–equivalent program graph IR that is faithful to the source code, and is solver-ready, forming the foundation for BMC and loop-condition encodings.

For each variable $x$ declared in the \texttt{VAR} section, we register a corresponding symbol in the \toolsmt AST, together with its type and domain constraints (which in \nusmv, we support Booleans and bounded integers).
Transition relations are handled by parsing the state updates of \nusmv into disjunctions of guard and update pairs (corresponding to the \conditions and \effects functions in a program graph). 
In addition, our parser supports predicates over states, which are defined as expressions evaluated at the current time step.
Here, we present the selected rules for parsing the initial conditions and transition blocks of a \nusmv model, where $x$ is a variable $e \in \effects$, $i \geq 0$ and $g \in \conditions$: 
% and the subscribed index represents the program locations:\borzoo{I don't understand the last}
\[
\begin{aligned}
% \llbracket \texttt{boolean} \rrbracket &= \mathsf{Bool} \qquad \\
% \llbracket \texttt{1..n} \rrbracket &= \mathsf{Int} \land 1\le x \le n\\
\llbracket \texttt{init}(x){:=}e \rrbracket &\triangleq  x_0 = \llbracket e \rrbracket_0 \\
\llbracket \texttt{next}(x){:=}e \rrbracket &\triangleq x_{i+1} = \llbracket e \rrbracket_i \\
% \llbracket \texttt{case } g_1{:}e_1;\cdots; \texttt{TRUE{:}}e_k\;\texttt{esac} \rrbracket_i
  % &= \bigvee_{j=1}^k \Big(\,\llbracket g_j \rrbracket_i \land \big(\llbracket x \rrbracket_{i+1} = \llbracket e_j \rrbracket_i\big)\,\Big)\\
\llbracket \texttt{DEFINE } \textit{pred} := e \rrbracket_i
  &\triangleq \textit{pred}_i \leftrightarrow \llbracket e \rrbracket_i \\
\llbracket e_1 \bowtie e_2 \rrbracket_i
  &\triangleq \big(\llbracket e_1 \rrbracket_i \;\bowtie\; \llbracket e_2 \rrbracket_i\big) \text{ where } \bowtie \in \{=,\neq,<,\le,>,\ge\}\\
\llbracket \land,\lor,\lnot,\rightarrow \rrbracket &\text{ map pointwise to \toolsmt Boolean connectives}\\
\llbracket +,-,*,/, \textit{mod } \rrbracket &\text{ map pointwise to \toolsmt arithmetic operations}
% \llbracket e_1 + e_2 \rrbracket_i &= \llbracket e_1 \rrbracket_i + \llbracket e_2 \rrbracket_i \quad \\
% \llbracket e_1 - e_2 \rrbracket_i &= \llbracket e_1 \rrbracket_i - \llbracket e_2 \rrbracket_i
\end{aligned}
\]
%
% \paragraph{Case Encoding.} 
For guarded updates expressed as a \texttt{case} statement in \nusmv, we interpret the effect as a disjunction of guarded assignments. 
Formally, for
\[
\texttt{next}(x) := \texttt{case } g_1 : e_1;\; g_2 : e_2;\; \cdots;\; \texttt{TRUE} : e_k \;\texttt{esac},
\]
the encoding at step $i$ requires
\[
x_{i+1} = 
\begin{cases}
\llbracket e_1 \rrbracket_i & \text{if } \llbracket g_1 \rrbracket_i \text{ holds}, \\
\llbracket e_2 \rrbracket_i & \text{if } \llbracket g_2 \rrbracket_i \text{ holds}, \\
\;\;\vdots & \\
\llbracket e_k \rrbracket_i & \text{otherwise (i.e., the `TRUE' case in \nusmv)}.
\end{cases}
\]
% Equivalently, in logical form we encode this as
% \[
% \bigvee_{j=1}^k \left( \llbracket g_j \rrbracket_i \;\land\; x_{i+1} = \llbracket e_j \rrbracket_i \right),
% \]
% which ensures that exactly one guarded effect determines the value of $x$ in the next step.
% {\color{red}
% Notice that, the locations of a program graph are determined by the current valuation of its variables.
% For clarity, we use $\llbracket v \rrbracket_i$ to denote the value of a variable $v$ at step $i$; $\llbracket g \rrbracket_i$ to denote the evaluation of all variables $v_1, v_2, \ldots$ occurring in a guard $g$ at step $i$; and $\llbracket e \rrbracket_i$ to denote the evaluation of all variables $v_1, v_2, \ldots$ occurring in an effect $e$ at step $i$.
% }
%
%Essentially, our parser translates \nusmv models into a structured Z3 AST (i.e., a SMT-LIB–equivalent representation), instantiating a well-formed program graph as the intermediate representation. 

% $j$ is the size of case enumeration defined in a \nusmv transition block. 
%
% and $i$ is a index indicator, which will later be used in the BMC unrolling. 

\subsection{Translating Verilog to IR}

In order to obtain an IR from \verilog, we rely on the \texttt{write\_smt2} 
command in \Yosys, which translates the design into an \toolsmt AST.
From this generated SMT-LIB encoding, we map the design to our IR by transforming the SMT-LIB into a variable-based transition system, decomposing the branching behavior into guard conditions, and then parsing it into our IR.

The \Yosys representation encodes the system as a state-wise transition relation, with the state modeled as an SMT-LIB datatype encompassing all program variables. 
The resulting model introduces variables of type \texttt{BitVec} and \texttt{Bool}, directly corresponding to those in the original design. 
It also defines several special-purpose functions: extractor functions for accessing individual variables, an initialization function that constrains the initial state, and a transition function relating a state $s$ to its successor $s’$. 
Since this output specifies the transition relation by representing the entire state as a datatype, we must first deconstruct the datatype and replace it with declared constants for each variable.
Additionally, we simplify the helper functions to accept only the necessary variables as parameters, and replace the extractor functions with the variables directly.

With this transformed SMT-LIB model, we now parse it into \toolsmt's \Rust wrapper.
Additional steps are taken during this process to memorize repeated structures in the tree caused 
by repeated helper function calls in some cases.
We create the transitions by recursively in-lining each helper function as it is encountered, and deconstruct each \texttt{(ite (condition) (true) (false))} into two transitions analogous to the method in Section~\ref{subsec:NuSMv2IR}.
%

%Essentially, the pre-post conditions for IR as defined as constants: \texttt{(declare-const |step\_0| |main\_s|)} for the initial state,
%and \texttt{(declare-const |step\_i| |main\_s|)}
%represents the unrolling on the $i$-th step.
%
%That is, the initial state must satisfy the initialization conditions specified in the \verilog design, and each subsequent state is derived by successively applying the transition function to its predecessor until all states are constrained accordingly, e.g.,
%\[
%\texttt{(assert (and (|main\_i| |step\_0|) (|main\_t| step\_0 %step\_1)\dots))}
%\]
%
%This enables us to parse \verilog models directly into IR\textcolor{red}{, enabling subsequent satisfiability checking with the chosen SMT/QBF backend}.
%
%To access valuations of variables relevant to the formula, we introduce constants of the appropriate types (i.e., \texttt{BitVec} or \texttt{Bool}), supporting construction of the fully unrolled model later in the BMC encoding.

% \newpage
\subsection{From IR to Bounded Model Checking Encoding}

The final stage bridges our IR with the SAT/QBF-based BMC unrolling, introduced in~\cite{hsb21,hbfs23}. 
Let $\varphi$ be a \HyperLTL formula of the form:
\[
\varphi = \quant_A\pi_A.\quant_B\pi_B \ldots \quant_Z\pi_Z.\psi, \qquad \quant \in \{\forall, \exists\},
\]
and let the set of IRs (i.e., ASTs) as $\pgs = \langle \programgraph_A, \programgraph_B, \dots, \programgraph_Z \rangle$.
% where each $\programgraph$ is the transition system induced by the program graph~\cite{baier2008principles} 
% (our IR), one per quantifier.\borzoo{Change to $\mathcal{P}$?}
%
The encoding of the \HyperLTL BMC problem with bound $k$ is the following:
\[
\qbf{\pgs, \varphi}^*_k = \quant_A\overline{x_A}.\quant_B\overline{x_B}.\cdots\quant_Z\overline{x_Z}.~ \Big( \qbf{\programgraph_A}_k 
\circ_A \qbf{\programgraph_B}_k \circ_B \cdots \qbf{\programgraph_Z}_k \circ_Z 
\QBFany{0}{k}{\psi} \Big),
\]
where $\qbf{\programgraph}_k$ is the $k$-step unrolling of a program graph $\programgraph$(i.e., enumerating concrete states according to the variable evaluations on $\conditions$ and $\effects$ functions),
$\QBFany{0}{k}{\psi}$ is the inner LTL formula unrolled for $k$ steps with the choice of semantics $*  \in \{\textit{pes},\textit{opt},\textit{hpes},\textit{hopt}\}$,
,
$\circ_j = \wedge$ if $\quant_j = \exists$, and
$\circ_j =\ \rightarrow$ if $\quant_j = \forall$, for each trace variable $j$ of $\varphi$.
For example, to perform bug-hunting for $\varphi_{\GNI}$, the negation $\neg\varphi_{\GNI}$ is of the form $\exists\exists\forall$, and build the encoding:
\begin{equation*}
	\qbf{\pgs, \neg \varphi_{\GNI}}_k = 
	\exists \overline{x_A}. \exists \overline{x_B}. \forall \overline{x_C}.
	\qbf{\programgraph_A}_k \wedge 
	\big(\qbf{\programgraph_B}_k \wedge 
	\big(\qbf{\programgraph_C}_k \rightarrow \qbf{\neg \psi}_k\big)\big).
\end{equation*}
From the \toolsmt AST–based IR in~\Cref{sec:IR}, we systematically derive the BMC encoding in either SMT or QBF, which the resulting SAT/UNSAT outcome is interpreted according to the chosen bounded semantics to determine the model checking result (see~\Cref{sec:spec_semantics}).

For \AHLTL, we adopt the encoding proposed 
in~\cite{hbfs23}, which involves additional quantifiers over \emph{trajectory variables} $\tau_a, \tau_b, \ldots, \tau_z$ for
parameterizing the \emph{position encoding} $\varphi_{pos}$.
This encoding introduces quantified sets of position variables $\overline{t_a}, \overline{t_b}, \ldots, \overline{t_z}$, together with auxiliary variables $\overline{\textit{pos}}$ and $\overline{\textit{off}}$ (for encoding trace positions while avoiding any trace to ``fall off'' from its bound), enabling the formalization of asynchronous alignments between traces (see~\cite{hbfs23} for details).
This mechanism allows different traces to advance at different speeds, thereby capturing asynchronous alignments that are not possible under the standard synchronous semantics. 
% The resulting encoding extends the traditional \HyperLTL BMC formulation by coupling trace quantifiers with trajectory quantifiers, ensuring that properties can be checked even when traces do not evolve in lock-step. 
The full encoding is given as follows:
\begin{align*}
\llbracket \pgs, \varphi \rrbracket_{k,m} =
\quant_A \overline{x_A}. \quant_B \overline{x_B}. \cdots \quant_Z \overline{x_Z}. \;
\quant_a \overline{t_a}. &\quant_b \overline{t_b}. \cdots \quant_z \overline{t_z}.~ \;
\exists \overline{\textit{pos}}. \; \exists \overline{\textit{off}}. \;\\
&\big( \llbracket \programgraph \rrbracket_k {\circ A} \cdots \llbracket \programgraph \rrbracket_k {\circ Z} \;
(\varphi_{pos} \wedge enc(\psi)) \big).
\end{align*}

% \tzuhan{think again, perhpas we don't need to explain this into this much details.}
% \paragraph{Transition system.}
% % We consider a model as the formal framework \emph{Kripke structure}.
% Let $AP$ be a finite set of \emph{atomic propositions} and $\Sigma = 2^{AP}$ be the
% \emph{alphabet}. 
% %
% A Kripke structure is a tuple
% $K = \langle S, S_{0}, \delta, L\rangle$, with a set of
% states $S$, a set of \emph{initial states} $S_{0} \subseteq S$, a
% \emph{transition relation} $\delta \subseteq S \times S$, and a
% \emph{labeling function} $L: S \rightarrow \Sigma$.
% %
% A \emph{path} of $K$ is an infinite sequence of states
% $s_0 s_1\ldots \in S^{\omega}$ such that $s_0 \in S_{0}$ and, for
% all $i \ge 0$, $(s_i, s_{i+1}) \in \delta$, 
% and the corresponding trace is a sequence $t(0)t(1)t(2)\cdots \in \Sigma^{\omega}$ with
% $t(i) = L(s(i))$ for all $i \ge 0$.
% %
% We write $\mathit{Traces}(K)$ as a shorthand for the set of traces of $K$ that
% start in all $s \in S_{0}$.

\subsection{From IR to Loop Conditions} 
The encoding in~\cite{hssb23} requires \emph{explicit states} in order to establish simulation relation over traces of programs with loops. 
To this end, we enumerate the \toolsmt AST IR to extract all reachable states along with the transition relation. 
While our IR in principle accommodates infinite state spaces, the simulation-based encoding assumes the state space is finite.
\begin{table}[t]
\centering
\caption{Overview of the required SAT constraints for loop condition encodings of $\aesim$ and $\easim$ over transition systems $\tupleof{K_P, K_Q}$~\cite{hssb23}.}
\scalebox{0.8}{
\begin{tabular}{m{.8cm}m{2cm}m{5.4cm}@{\hspace{6mm}}m{5.8cm}}
\toprule
\# & \textbf{Constraint} & \textbf{$\aesim$} & \textbf{$\easim$} \\
\midrule
\textbf{(c1)} & Reachable states 
% &\multicolumn{2}{c}{ All states in both $K_P$ and $K_Q$ must encoding only reachable states }\\
& All states in both $K_P$ and $K_Q$ must encoding only reachable states
& Same as $\aesim$ \\
\midrule
\textbf{(c2)} & Exhaustive exploration 
& $K_P$ must be fully explored; every distinct state is represented 
& $K_Q$ must be fully explored; every distinct state is represented  \\
\midrule
\textbf{(c3)} & Initial-state simulation 
& \emph{Every} initial states of $K_P$ must simulates \emph{some} initial state of $K_Q$ 
& There \emph{exists} an initial state of $K_P$ that simulates \emph{every} initial state of $K_Q$ \\
\midrule
\textbf{(c4)} & Successor preservation 
& \emph{Every} successor in $K_P$ must match a successor in $K_Q$
& A successor in $K_P$ must match \emph{every} successors in $K_Q$, including jump-backs \\
\midrule
\textbf{(c5)} & Predicate consistency 
& Relational predicates must be preserved across all simulated pairs 
& Same as $\aesim$ \\
\bottomrule
\end{tabular}
}
\vspace{-5mm}
\label{tab:sim-encodings}
\end{table}
% The encoding of $\easim$ ensures that one system $K_P$ (quantified by $\exists$) can simulate another system $K_Q$ (quantified by $\forall$) by enforcing the following conditions:
% \begin{itemize}
%   \item \textbf{Reachable states:} only legal encodings of states are considered. 
%   \item \textbf{Exhaustive exploration:} all distinct states of $K_Q$ must be represented. 
%   \item \textbf{Initial-state simulation:} the initial state of $K_P$ must simulate all initial states of $K_Q$. 
%   \item \textbf{Successor preservation:} every transition in $K_Q$ is matched by a corresponding transition in $K_P$, including possible loop-backs. 
%   \item \textbf{Predicate consistency:} relational predicates over labels of states must be preserved under simulation. 
% \end{itemize}
% Together, these constraints yield a complete SAT encoding for deciding whether an EA-simulation exists from $K_P$ to $K_Q$.

Table \ref{tab:sim-encodings} summarizes the loop condition encodings for $\aesim$ (for $\forall\exists~ \psi$) and $\easim$ for $\exists\forall~\psi$). 
Essentially, given two models $K_P$ and $K_Q$ (one per trace quantifier), both encodings require reachable states to be respected and predicates to remain consistent across simulated pairs (i.e., constraints \textbf{(c1)} and \textbf{(c5)}). 
The key differences arise in the treatment of exploration, initial states, and successor preservation (i.e., constraints \textbf{(c2)}, \textbf{(c3)}, and \textbf{(c5)}, respectively). 
% For $\mathsf{SIM}_{\forall\exists}$, all states of $K_P$ must be fully explored, and every initial state of $K_P$ must simulate some initial state of $K_Q$, with every successor in $K_P$ matching a successor in $K_Q$. 
%
% In contrast, $\mathsf{SIM}_{\exists\forall}$ demands exhaustive exploration of $K_Q$, requires the existence of at least one initial state of $K_P$ that simulates all initial states of $K_Q$, and strengthens successor preservation by requiring each successor in $K_P$ to match every successor in $K_Q$, including possible jump-backs (i.e., step into a state that is already in the simulation). 
%
This distinction reflects the dual quantifier structure of the two conditions, balancing universal and existential quantifiers in the simulation relation.
Detailed SAT encodings are available in~\cite{hssb23}.

% \Cref{tab:sim-encodings} presents the overview of the SAT constraints of $\aesim$ (for $\forall\exists~ \psi$) and $\easim$ for $\exists\forall~\psi$). 
% Due to space, the full encodings are deferred to the appendix.
% In essence, given two models $K_P$ and $K_Q$ (one per trace quantifier), the encoding verifies whether an infinite trace in one model (i.e., a lasso path) can simulate the behavior of another lasso path in a different model. 
% %
% This allows infinite behaviors of  (i.e., verify a $\mathbf{G} \psi$ formula) to be captured through finite exploration in a simple and concise manner. 
% \newpage

%% file: experiments.tex
\vspace{-2mm}
\section{Empirical Evaluation}
\label{sec:eval}
%\vspace{-2mm}

We extensively evaluated \HyperQB on a comprehensive suite of benchmarks spanning all major dimensions: 
(1) different logics (\HyperLTL and \AHLTL), (2) input languages (\nusmv and \verilog), and (3) direct comparisons with the model checker \AutoHyper~\cite{bf23} and the earlier prototype of \HyperQB, where applicable.
% \tzuhan{"infinite" is iff we have C cases or NuXMV with unbounded ints, check back.}
%
We are also significantly expanding our benchmark suite and case studies, creating a wealth of 
examples for the research community. The detailed description of all benchmarks can be found at: 
\url{https://hyperqb.github.io/case-studies}.
For clarity, in this section, we write $\text{HQ2.0}_{SMT}$ to denote the {\em total} model checking time for \HyperQB with \toolsmt as the backend solver, and $\text{HQ2.0}_{QBF}$ to denote the configuration using \quabs. 
%
% The detailed descriptions of each case presented in this section can be found in our official website~\footnote{https://hyperqb.github.io/}.
%
%This comprehensive coverage highlights the versatility and robustness of our framework across a wide spectrum of verification scenarios.
%
All experiments are run on an MacBook Pro with Apple M1 Max chip and 64 GB of memory.
% \tzuhan{Fill XXX when complete.}

% \paragraph{HyperLTL cases} \tzuhan{TODO.}
\begin{table}[b]
\vspace{-5mm}
    \caption{Summary of results for HyperLTL benchmarks from~\cite{hsb21}. Here, $\quant$ denotes the form of quantifiers in the given formula, $k$ denotes the unrolling bound; `Encoding' is the time to construct the unrolled IR; `AH' refers to \AutoHyper~\cite{bf23} run time; `verification' reports only SAT/UNSAT result, a dash `-' indicates that no witness should be returned for the given case, since the result is either verified ($\cmark$) with a leading $\forall$, or falsified ($\xmark$) for a leading $\exists$. When the backend solver is QBF, a counterexample is always given when the case applies.}
    \centering
    \vspace{1mm}
        \scalebox{0.87}{
    \input{tables/table_hltl1}

    }
    \vspace{-4mm}
    \label{tab:result-hltl}
\end{table}

\vspace{-3mm}
\subsection{HyperLTL Benchmarks from~\cite{hsb21} (\Cref{tab:result-hltl})}
\label{sec:oldcases}

% \tzuhan{remove the $*$ and \# when completed.}
We begin our evaluation with the benchmark suite originally appeared in~\cite{hsb21} (see~\Cref{tab:result-hltl}), including hyperproperties expressing symmetry, linearizability, non-interference, fairness, and robotic path planning. 
%
%Importantly, it also includes instances with large state spaces (i.e., path planning cases), thereby serving as a challenging testbed for assessing the scalability and robustness of model checkers.
%
Our first observation is that \ourtool delivers a substantial improvement over the original prototype (denoted as HQ1.0), independent of whether the backend relies on an SMT or QBF solver. 
The optimizations in the new version (most notably, the adoption of \toolsmt ASTs and the \Rust implementation) yield significant performance gains across the entire benchmark suite. 

\begin{table}[b!]
\vspace{-5mm}
\caption{Results for new HyperLTL benchmarks. $\quant$ denotes the form of quantifiers in the given formula, $k$ denotes the unrolling bound; `Encoding' is the time to construct the BMC encoding; `AH' refers to \AutoHyper run time; verification reports only SAT/UNSAT results, a dash `-' indicates that no witness should be returned for the given case, since the result is either verified ($\cmark$) with a leading $\forall$, or falsified ($\xmark$) for a leading $\exists$. When the backend solver is QBF, a counterexample is always given when the case applies.}
    \vspace{1mm}
    \centering
        \scalebox{0.88}{
    \input{tables/table_hltl2}

    }
    \vspace{-4mm}
    \label{tab:result-hltl2}
\end{table}

For comparison with \AutoHyper, since the same benchmarks were used in~\cite{bf23} (which the authors compared with our earlier prototype), we re-ran all experiments and observe that \ourtool significantly outperforms \AutoHyper in the vast majority of cases, demonstrating clear advantages in both expressiveness and efficiency. Several numbers differ from those originally reported in~\cite{bf23}. We communicated this to the authors and they confirmed that this discrepancy is due to version updates made after publication. For fairness, we report results using the most up-to-date version of \AutoHyper as of September 1, 2025.

\vspace{-2mm}
\subsection{New Benchmarks on HyperLTL (\Cref{tab:result-hltl2})}
\label{sec:newcases}
\vspace{-2mm}

In addition to the original benchmark suite from~\cite{hsb21}, we extend our evaluation with 29 new case studies to further assess the efficiency and effectiveness of \HyperQB.
% \tzuhan{update XXX when completed.}
%
These new benchmarks primarily focus on two aspects: (1) formulas with more than two quantifiers, which increase the computational complexity of verification, and (2) models exhibiting strong non-determinism, which challenge the scalability and robustness of the underlying encodings.

% \tzuhan{remove the $*$ and \# when completed.}
The results, summarized in~\Cref{tab:result-hltl2}, demonstrate that \HyperQB significantly outperforms \AutoHyper across the vast majority of the new benchmarks. 
% \tzuhan{P.S the blank parts on this table are which QBF got TO, but based on other cases, it shouldn't. I plan to investigate and try those cases again after the write-up is ready, before claiming "TO" on those blanks.}
%
In particular, for formulas with more than two quantifiers (e.g., MapSynth 2 with five quantifiers $\exists\forall\forall\exists\exists$), \AutoHyper incurs heavy overhead to constructing automata complementation, which is its core technique to handle quantifier alternation. 
Likewise, in cases with substantial input or transition non-determinism, \AutoHyper’s reliance on language inclusion checking leads to significant performance overhead (e.g., NDET v2, NDET v3, and IQueue). 
By contrast, these factors have little impact on \HyperQB, highlighting the robustness of our SMT-based approach. 
This illustrates not only the scalability of \HyperQB to complex quantifier structures and nondeterministic models, but also its versatility as a general framework for hyperproperty verification.

\begin{table}[b!]
\vspace{-2mm}
    \caption{Summary of results for \AHLTL benchmarks from~\cite{hbfs23}, where $\quant$ denotes the form of quantifiers in the given \AHLTL formula (we use $\{\textsf{A},\textsf{E}\}$ to specify trajectory quantifiers from path quantifiers (i.e., $\{\forall, \exists\}$), $k$ is the unrolling bound, $M$ is the trajectory bound, and `Encoding' is the time to construct the BMC encoding.}
    \vspace{1mm}
    \centering
    \scalebox{0.9}{
    \input{tables/table_ahltl}
    }
    % \vspace{-4mm}
    \label{tab:result-ahltl}
\end{table}

\vspace{-2mm}
\subsection{\AHLTL Benchmarks from~\cite{hbfs23} (\Cref{tab:result-ahltl})}
\label{sec:ahltlcases} 
\vspace{-2mm}

We now turn to benchmarks involving \emph{asynchronous} hyperproperties, drawn from~\cite{hbfs23}. 
These include challenging scenarios such as information-flow in concurrent programs, speculative execution, compiler optimizations, and cache-based timing attacks (details of properties in~\cite{hbfs23}). 
Presented in~\Cref{tab:result-ahltl}, where all cases require \AHLTL formulas because the relevant traces do not proceed in lock-step. 
We emphasize that \AutoHyper cannot deal with asynchronous hyperproperties, so no comparison is included in~\Cref{tab:result-ahltl}. 
%
%Because these reductions are non-trivial and not automated, reproducing the \AHLTL cases for comparison is infeasible.
%
\HyperQB constitutes a breakthrough: to the best of our knowledge, it is the \emph{first and only} tool to offer automated model checking for \AHLTL. 
%
%This enables users to directly verify asynchronous hyperproperties without manual encodings or reductions, marking a significant advance in the state-of-the-art. 
%
%This capability expands the practical reach of hyperproperty verification to a much broader class of real-world systems, including those whose security or correctness inherently depends on reasoning about traces that advance at different speeds, such as secure compilation~\cite{pg17}.

Our experimental results in~\ref{tab:result-ahltl} show that \ourtool consistently outperforms the hard-coded implementation in~\cite{hbfs23}, regardless of the chosen backend solver (SMT or QBF).
Notably, in several benchmarks the QBF backend exhibits an advantage: this is because the \AHLTL position encoding, originally proposed in~\cite{hbfs23}, is fully Booleanized and thus naturally aligned with QBF solvers.
By contrast, using SMT backends on such encodings may introduce unnecessary overhead due to the additional reasoning over theories.
Nevertheless, across the benchmark suite \ourtool has a significantly reduced solving time compared to~\cite{hbfs23}, demonstrating the robustness of our solver-based design.

\begin{table}[b!]
	  \begin{minipage}{.52\linewidth}
    \caption{Performance results: encoding, solving, total time, and comparison with~\cite{hssb23}.}
    \vspace{1mm}
    \centering
    \scalebox{.8}{
    \input{tables/table_loop}
}
    \vspace{-5mm}
\label{tab:performance}
\end{minipage}
%\end{table}
%%%%%%%%%%%%%%%  (separator to avoid conflict ) %%%%%%%%%%%%%%%
%\begin{table}[ht!]
\hfill
	  \begin{minipage}{.35\linewidth}
\caption{Performance on selected case studies in \verilog.}
	\centering
    \scalebox{.8}{
    			\begin{tabular}{|l|c|c|c|c|c|}
				\hline
				Case Study & $k$ & Parse & Encode & HQ2.0 \\ \hline \hline
				$LED_{EA_{true}}$ & 101 & 0.03 & 1.08 & 0.217 \\ \hline
				$LED_{AE_{true}}$ & 10 & 0.03 & 0.07 & 119.3 \\ \hline
				$LED_{EE_{false}}$ & 101 & 0.03 & 0.17 & 0.002\\ \hline
				$LED_{EE_{true}}$ & 101 & 0.03 & 0.16 & 0.016 \\ \hline
				SPI & 8 & 0.27 & 0.05 & 0.016\\ \hline
				FPU2 & 8 & 0.16 & 0.14 & 0.473 \\ \hline

		\end{tabular}
}
\label{tab:results_verilog}

\end{minipage}
\end{table}

\vspace{-3mm}
\subsection{Benchmarks of Programs with Loops 
from~\cite{hssb23} (\Cref{tab:performance})}

Recall that the benchmarks in~\Cref{sec:oldcases,sec:newcases,sec:ahltlcases} are all loop-free. 
Reasoning about infinite semantics in BMC, however, has long been a major challenge, leaving a significant gap in prior approaches. \ourtool closes this gap by implementing the loop conditions identified in~\cite{hssb23} and it can seamlessly and efficiently verify global properties (i.e., \HyperLTL formulas with the $\mathbf{G}$ operator) in the presence of loops.

We construct the loop-conditions encoding directly from IR using distinct strategies tuned for performance. For the $\forall\exists$ case, we explicitly enumerate states and build a transition matrix. Because the $\forall\exists$ encoding of~\cite{hssb23} relies on pairwise simulation, having concrete state values and an explicit transition matrix makes constraint checking substantially easier for the solver. By contrast, for the $\exists\forall$ case, we use an implicit-state formulation that lets the solver assign state values subject to the initial constraints and the transition relation. Here, the encoding must find a lasso in the first model that simulates all paths of the second model. Constructing such a lasso over explicit states would require an exhaustive exploration of paths and leads to a heavy encoding, whereas constrained implicit states allow the solver to build the lasso on the fly.

\Cref{tab:loopcond} shows that our new implementation within \ourtool for all the benchmarks significantly outperforms those reported in~\cite{hssb23}.

% Notes-Observations:
% \begin{enumerate}
%     \item HyperQB is more efficient in solving cases with large state spaces (e.g., MapSynth2, EMM ABA, etc.).
%     \item HyperQB synthesizes counterexamples without significant overhead (e.g., NDET v2, NDET v3, Bank v1, BufferOD, IQueue, etc.).
%     \item HyperQB handles input non-determinism and transition non-determinism better (e.g., NDET v2, NDET v3, Keypad, etc.).
% \end{enumerate}

\vspace{-3mm}
\subsection{Hardware Security Benchmarks in Verilog (\Cref{tab:results_verilog})}

A subset of our evaluation focuses on hardware-oriented designs written in \verilog.
These case studies are drawn from the benchmarks used in~\cite{Gleissenthall2019iodine, bcbfs21}.
These cases cover constant time execution and observational determinism, as hardware-level implementations often expose subtle timing and information-flow vulnerabilities.

The first case study is a minimal blinking LED design that includes multiple internal states within the on phase. 
We evaluate formulas that assert observational determinism, requiring that all execution paths produce the same external LED behavior regardless of internal state.

One representative case study for constant time execution involves the design of a Floating-Point Unit (FPU) which is a component in a CPU designed for fast, precise calculations with floating-point numbers.
We selected the division module of this unit for verification.
These benchmarks are especially useful for studying time side-channel attacks, as even small differences in control paths can manifest as observable variations in execution time.
We measure this execution time in terms of observable registers and how they are updated in relation to clock cycles.
In our encoding, the \texttt{unsat} outcome is driven by exception paths that assert status signals earlier than the nominal datapath completes. 
The module also exhibits data-dependent timing for nominal values as well, producing results after varying numbers of cycles.
These early exits create timing differences, allowing an attacker to infer information from when the module updates status before the expected cycle count of the operation.
In comparison with the results in ~\cite{Gleissenthall2019iodine}, for the purpose of bug discovery \ourtool outperforms their result when not restricting the operand values.

From~\cite{bcbfs21}, we focus on the SPI bus case study which is a representative example of verifying observational determinism in hardware designs.
The SPI bus is a widely used serial communication protocol commonly used for communication with peripheral devices.
In this case, we want to ensure that secret of high-level inputs do not lead to observable differences in the behavior of the system from the perspective of a low-level observer.
For the SPI bus this means verifying that any two executions with the same initial state and output must also agree on their bus configurations. The result of SAT indicates that this model satisfies

Together, the LED, SPI, and FPU cases demonstrate \HyperQB's ability to handle various hardware security concerns.
The FPU case also demonstrates \HyperQB's ability to scale beyond toy examples and handle real-world hardware modules, providing strong evidence of its practical applicability.

%%%%%%%%%%%%%%%  (separator to avoid conflict ) %%%%%%%%%%%%%%%

% \paragraph{Complexity analysis (crafted examples)} \tzuhan{TBD.}

%% file: tables/table_hltl1.tex
\begin{tabular}{|c |l|c|c||c || c|c || c|c || c| c| c }
    \cline{6-11} 
    \multicolumn{5}{c}{ }
    &\multicolumn{2}{|c|}{\bf Verification Only}
    &\multicolumn{4}{|c|}{\bf Verification + Witness} \\
    \cline{1-11}
    & {\bf Benchmark} & $\quant$ & $k$ & {\bf Encoding} & {\bf HQ2.0$_{\mathbf{SMT}}$} & AH & {\bf HQ2.0$_{\mathbf{SMT}}$} & AH & HQ1.0 & {\bf HQ2.0}$_{\mathbf{QBF}}$  \\ \cline{1-11}\cline{1-11}
    \multirow{18}{*}{\rotatebox{90}{\bf Benchmarks from~\cite{hsb21}}} 
    & Bakery3 & $\forall\exists$ & $10$  & 0.01 & \better{0.09} & 0.72 & \bettercex{0.04} & 0.67 & \ocamlnum{6.31} & \rustnum{0.20} & $\xmark$ \\ \cline{3-11}
    & Bakery7 & $\forall\exists$ & $10$  & 0.05 & \better{0.16} & 3.80 & \bettercex{0.06} & 1.87 & \ocamlnum{19.73} & \rustnum{3.83} & $\xmark$ \\ \cline{3-11}
    & Bakery9 & $\forall\exists$ & $10$  & 22.17 & \better{0.17} & 6.33 & \bettercex{0.07} & 5.41 & \ocamlnum{TO} & \rustnum{TO} & $\xmark$  \\ \cline{3-11}
    & Bakery11 & $\forall\exists$ & $10$  & 0.03 & \better{0.16} & 34.80 & \bettercex{0.07} & 33.47 & \ocamlnum{TO} & \rustnum{TO} & $\xmark$   \\ \cline{2-11}
    & SNARK1 & $\forall\exists$ & $18$  & 0.10 & \better{1.87} & 7.80 & \bettercex{0.93} & 6.38 & \ocamlnum{72.37} & \rustnum{2.01} & $\xmark$  \\ \cline{2-11}
    % & SNARK2 & ? & ? & ? & ? & ? & ? & ?  & ? \\ \cline{3-11}
    & NI$_{Correct}$ & $\forall\exists$ & $50$  & 0.11 & 1.43 & \better{0.62} & - & - & \ocamlnum{1.40} & \rustnum{1.53} & $\cmark$ \\ \cline{3-11}
    & NI$_{Inorrect}$ & $\forall\exists$ & $50$  & 0.11 & \better{1.38} & 2.59 & 7.20 & \bettercex{1.91} & \ocamlnum{5.24} &\rustnum{1.52} & $\xmark$ \\ \cline{2-11}
    & NRP$_{Correct}$ & $\exists\forall$ & $15$  & 0.01 & \better{0.16} & 0.70 & \bettercex{0.14} & 3.60 & \ocamlnum{0.96} & \rustnum{0.85} & $\cmark$ \\ \cline{3-11}
    & NRP$_{Incorrect}$ & $\exists\forall$ & $15$ & 0.01  & \better{0.13} & 1.21 & - & - & \ocamlnum{1.05} & \rustnum{0.83} & $\xmark$ \\ \cline{2-11}
    & Robust $10^2$ & $\forall\exists$ & $20$ & 0.04 & \forq{\better{0.38}} & \forq{5.11} & - & -& \ocamlnum{3.01} & \rustnum{1.51} & $\cmark$ \\ \cline{3-11}
    & Robust $20^2$ & $\forall\exists$ & $40$  & 0.11 & \forq{\better{4.05}} & \forq{22.47} & - & - & \ocamlnum{10.05} & \rustnum{5.66} & $\cmark$  \\ \cline{3-11}
    & Robust $40^2$ & $\forall\exists$ & $80$  & 0.97 & \forq{\better{9.13}} & \forq{209.87} & - & - & \ocamlnum{110.60} & \rustnum{19.67} & $\cmark$  \\ \cline{3-11}
    & Robust $60^2$ & $\forall\exists$ & $120$  & 6.78 & \better{1233.39} & \forq{TO} & - & - & \ocamlnum{478.42} & \rustnum{129.88} & $\cmark$  \\ \cline{2-11}
    & SP $10^2$ & $\exists\forall$ & $20$ & 0.08 & \better{0.67} & 5.23 & \bettercex{1.69} & 2.68 & \ocamlnum{6.53} & \rustnum{2.01} & $\cmark$  \\ \cline{3-11}
    & SP $20^2$ & $\exists\forall$ & $40$  & 0.96 & {32.75} & \better{13.61} & 885.91 & \bettercex{3.80} & \ocamlnum{110.48} & \rustnum{TO} & $\cmark$  \\ \cline{3-11}
    & SP $40^2$ & $\exists\forall$ & $80$   & 26.96 & \forq{TO} & \better{85.56} & TO & \bettercex{17.18} & \ocamlnum{TO} & \rustnum{TO} & $\cmark$ \\ \cline{3-11}
    & SP $60^2$ & $\exists\forall$ & $120$   & 3.56 & \forq{TO} & \better{41.24} & TO & \bettercex{41.35} & \ocamlnum{TO} & \rustnum{TO} & $\cmark$  \\ \cline{2-11}
    & Mutation & $\exists\forall$ & $10$   & 0.01 & \better{0.05} & 0.52 & \bettercex{0.03} & 0.36 & \ocamlnum{0.83} & \rustnum{0.85} & $\cmark$ \\ \cline{1-11}\cline{1-11}
\end{tabular}

%% file: tables/table_hltl2.tex
\begin{tabular}{|l|c| c ||c || c|c || c|c || c| c| c }
    \cline{5-10} 
    \multicolumn{4}{c}{ }
    &\multicolumn{2}{|c}{\bf Verification Only }
    &\multicolumn{4}{|c|}{\bf Verification + Witness}\\
    \cline{1-10}
     {\bf Benchmark} & $\quant$ & $k$ & {\bf Encoding} & {\bf HQ2.0$_{\mathbf{SMT}}$} & AH & {\bf HQ2.0$_{\mathbf{SMT}}$} & AH & HQ1.0 & {\bf HQ2.0}$_{\mathbf{QBF}}$  \\ \cline{1-10}\cline{1-10}
    %% 
    % \multirow{16}{*}{\rotatebox{90}{HyperQB 1.0 arXiv cases}} 
     Co-term & $\forall\forall$ & $10$  & 0.00 & \better{0.01} & 0.30 & - & - & \ocamlnum{1.65} & \rustnum{4.10} & $\cmark$ \\ \cline{1-10}
     Deniability & $\forall\exists\exists$ & $10$  & 0.01 & \better{0.15} & \forq{11.56} & - & - & \ocamlnum{105.23} & \rustnum{17.78} & $\cmark$  \\ \cline{1-10}
     Buffer$_{\mathit{OD}^{\mathit{c}}}$ & $\forall\forall$ & $10$  & 0.01 & \better{0.04} & 0.43 & \bettercex{0.04} & 0.85 & \ocamlnum{2.70} & \rustnum{0.88} & $\xmark$  \\ \cline{2-10}
     Buffer$_{\mathit{OD}^{\mathit{i}}}$ & $\forall\forall$ & $10$   & 0.01 & \better{0.04} & 0.68 & - & - & \ocamlnum{2.63} & \rustnum{0.82} & $\cmark$ \\ \cline{2-10}
     Buffer$_{\mathit{GMNI}^{\mathit{i}}}$ & $\forall\exists$ & $10$  & 0.01 & \better{0.09} & 2.69 & - & - & \ocamlnum{1.42} & \rustnum{0.81} & $\cmark$ \\ \cline{2-10}
     Buffer$_{\mathit{OD}^{\mathit{c}}}$ & $\forall\forall$ & $10$  & 0.00 & \better{0.03} & 0.56 & \bettercex{0.04} & 9.11 & \ocamlnum{1.24} & \rustnum{0.82}  & $\xmark$ \\ \cline{1-10}
     Nodet$_{\mathit{TINI}}$ & $\forall\exists$ & $10$   & 0.00 & \better{0.09} & \forq{1.85} & - & -  & \ocamlnum{2.31} & \rustnum{0.77} & $\cmark$ \\ \cline{2-10}
     Nodet$_{\mathit{TSNI}}$ & $\forall\exists$ & $10$  & 0.01 & \better{0.08} & \forq{1.83} & - & - & \ocamlnum{2.35} & \rustnum{0.78} & $\cmark$  \\ \cline{1-10}
     K-Safety & $\forall\forall$ & $10$   & 0.02 & \doublecheck{6.89} & \better{0.49} & - & - & \ocamlnum{2.84} & \rustnum{TO} & $\cmark$  \\ \cline{1-10}
     MapSynth1 & $\exists\forall\forall\exists\exists$ & $10$   & 0.01 & \better{0.07} & 0.50 & TO & \bettercex{0.34} & \ocamlnum{0.79}  &  \rustnum{0.06} & $\cmark$ \\ \cline{2-10}
     MapSynth2 & $\exists\forall\forall\exists\exists$ & $10$   & 0.00 & \better{30.27} & \forq{TO} & TO & TO & \ocamlnum{9.49} & \rustnum{TO} & $\cmark$  \\ \cline{1-10}
     TeamLTL$_{\text{v1}}$ & $\exists\exists\forall$ & $10$   & 0.01 & 1.39 & \better{0.36} & - & - & \ocamlnum{1.58} & \rustnum{0.92} & $\xmark$ \\ \cline{2-10}
     TeamLTL$_{\text{v2}}$ & $\exists\exists\forall$ & $20$   & 0.02 & 5.80 & \better{0.89} & - & - & \ocamlnum{60.08} & \rustnum{0.93} & $\xmark$  \\ \cline{1-10}
     NDET$_{\text{v1}}$ & $\forall\exists$ & $10$   & 0.00 & \better{0.03} & 0.31 & \bettercex{0.03} & 0.34 & \ocamlnum{0.69} & \rustnum{0.76} & $\xmark$ \\ \cline{2-10}
     NDET$_{\text{v2}}$ & $\forall\exists$ & $10$   & 0.00 & \better{0.04} & \forq{0.35} & \bettercex{0.03} & \forq{TO}  & \ocamlnum{0.73} & \rustnum{0.81} & $\xmark$  \\ \cline{2-10}
     NDET$_{\text{v3}}$ & $\forall\exists$ & $10$   & 0.00 & \better{0.04} & \forq{68.99} & \bettercex{0.03} & \forq{TO} & \ocamlnum{0.76} & \rustnum{0.84} & $\xmark$  \\ \cline{1-10}
    %%
    % \multirow{13}{*}{\rotatebox{90}{HyperQB 2.0 new cases}} 
     Bank$_{\text{v1}}$ & $\forall\forall\exists$ & $10$   & 0.01 & \better{0.10} & 0.53 & \bettercex{0.04} & 11.48 & \ocamlnum{51.45} & \rustnum{0.87} & $\xmark$  \\ \cline{2-10} 
     Bank$_{\text{v2}}$ & $\forall\forall\exists$ & $10$  & 0.01 & \better{0.06} & 0.32 & \bettercex{0.05} & 0.68 & \ocamlnum{2.68} & \rustnum{0.83} & $\xmark$  \\ \cline{2-10} 
     Bank$_{\text{v3}}$ & $\forall\forall\exists$ & $10$   & 0.01 & \better{0.07} & 0.33 & \bettercex{0.04} & 0.74 & \ocamlnum{2.16} & \rustnum{0.82} & $\xmark$ \\ \cline{1-10} 
     Constructor & $\forall\exists$ & $10$   & 0.01 & \better{0.08} & 0.58 & \bettercex{0.08} & 0.42 & \ocamlnum{1.77} & \rustnum{0.89} & $\cmark$   \\ \cline{1-10}
     Bidding$_{\text{v1}}$ & $\forall\forall$ & $10$   & 0.01 & \better{0.01} & 0.28 & - & - & \ocamlnum{0.73} & \rustnum{0.95} & $\cmark$  \\ \cline{2-10}
     Bidding$_{\text{v2}}$ & $\forall\forall$ & $10$   & 0.01 & \better{0.01} & 0.29 & - & - & \ocamlnum{0.86} & \rustnum{0.89} & $\cmark$ \\ \cline{2-10}
     Bidding$_{\text{v3}}$ & $\forall\forall$ & $10$   & 0.01 & \better{0.02} & 0.68 & - & - & \ocamlnum{30.61} & \rustnum{1.07} & $\cmark$ \\ \cline{2-10}
     Bidding$_{\text{v4}}$ & $\forall\forall$ & $10$   & 0.01 & \better{0.05} & 0.32 & \bettercex{0.03} & 2.81 & \ocamlnum{1.89} & \rustnum{0.95} & $\xmark$   \\ \cline{1-10}
     IQueue & $\forall\exists$ & $10$   & 0.06 & \better{0.73} & \forq{TO} & \bettercex{0.59} & \forq{TO} & \ocamlnum{68.06} &  \rustnum{0.80} & $\xmark$ \\ \cline{1-10}
     Keypad & $\exists\exists$ & $10$   & 0.01 & \better{0.02} & 14.57 & \bettercex{0.01} & 10.65 &  \ocamlnum{1.48} & \rustnum{0.65} & $\cmark$  \\ \cline{1-10}
     SimpleQueue & $\forall\forall$ & $10$   & 0.03 & \better{0.02} & 0.27 & \bettercex{0.02} & 0.22  & \ocamlnum{0.75} & \rustnum{0.26} & $\xmark$ \\ \cline{1-10} 
     EMM ABA & $\forall\exists$ & $10$  & 0.02 & \better{0.12} & \forq{72.52} & 0.07 & \forq{TO} & \ocamlnum{88.14} & \rustnum{0.15} & $\xmark$ \\ \cline{1-10} 
     Lazy List & $\forall\exists$ & $13$   & 0.05 & \better{0.40} & \forq{TO} & \bettercex{0.20} & \forq{TO} & \ocamlnum{TO} & \rustnum{0.43} & $\xmark$  \\ \cline{1-10}\cline{1-10}
\end{tabular}

%% file: tables/table_ahltl.tex
\newcommand{\Etau}{\textsf{E}}
\newcommand{\Atau}{\textsf{A}}
\begin{tabular}{|l|c|c|c|c|c|c||c|c}
    \cline{1-8}
    {\bf Benchmark} & $\quant$ & $k$ & $M$ & {\bf Encoding} & {\bf HQ2.0$_{\mathbf{SMT}}$} & \cite{hbfs23} & HQ2.0$_{QBF}$ \\ \cline{1-8}
    \cline{1-8}
    ACDB & $\forall\exists\Etau$ & $11$ & $22$   & 0.22 & \better{0.14} & 3.35 & \rustnum{0.79} & $\xmark$   \\ \cline{2-8}
    ACDB$_{nd}$ & $\forall\exists\Etau$ & 8 & 16   & 0.08 & \better{1.38} & 14.01 & \rustnum{2.20} & $\xmark$  \\ \cline{1-8}
    ConcLeak & $\forall\exists\Etau$ & 11 & 22   & 0.14 & \better{5.10} & 30.24 & \rustnum{8.13} & $\xmark$ \\ \cline{2-8}
    ConcLeak$_{nd}$ & $\forall\exists\Etau$ & 18 & 36   & 0.40 & TO & 911.72 & \better{44.22} & $\xmark$ \\ \cline{1-8}
    SpecExcu$_{V1}$ & $\forall\exists\Etau$ & 6 & 12    & 0.04  & \better{0.17} & 12.24 & \rustnum{0.65} & $\xmark$ \\ \cline{2-8}
    SpecExcu$_{V2}$ & $\forall\exists\Etau$ & 6 & 12    & 0.04 & \better{0.12} & 9.33 & \rustnum{0.52} & $\cmark$ \\ \cline{2-8}
    SpecExcu$_{V3}$ & $\forall\exists\Etau$ & 6 & 12    & 0.04 & \better{0.20} & 11.95 & \rustnum{0.71} & $\xmark$ \\ \cline{2-8}
    SpecExcu$_{V4}$ & $\forall\exists\Etau$ & 6 & 12    & 0.05 & \better{0.18} & 13.58 & \rustnum{0.66} & $\xmark$ \\ \cline{2-8}
    SpecExcu$_{V5}$ & $\forall\exists\Etau$ & 6 & 12    & 0.04 & \better{0.15} & 10.94 & \rustnum{0.62} & $\cmark$ \\ \cline{2-8}
    SpecExcu$_{V6}$ & $\forall\exists\Etau$ & 6 & 12   & 0.04 & \better{0.16} & 12.55 & \rustnum{0.77} & $\xmark$ \\ \cline{2-8}
    SpecExcu$_{V7}$ & $\forall\exists\Etau$ & 6 & 12    & 0.03 & \better{0.12} & 9.69 & \rustnum{0.54} & $\cmark$ \\ \cline{1-8}
    DBE & $\forall\forall\Etau$ & 4 & 8 & 0.01 & \better{0.04} & 0.98 & \rustnum{0.81} & $\cmark$ \\ \cline{2-8}
    DBE$_{ndet}$ & $\forall\forall\Etau$ & 13 & 26  & 0.07 & \better{0.21} & 11.77 & \rustnum{2.84} & $\cmark$ \\ \cline{2-8}
    DBE$_{ndet}^{bugs}$ & $\forall\forall\Etau$ & 13 & 26   & 0.07 & \better{0.20} & 3.90 & \rustnum{2.83} & $\xmark$ \\ \cline{1-8}
    LP & $\forall\forall\Etau$ & 22 & 44  & 0.20 & 933.21  & {3.90} & \better{2.54} & $\cmark$ \\ \cline{2-8}
    LP$_{ndet}$ & $\forall\forall\Etau$ & 17 & 34    & 0.11 & 415.21 & 59.38 & \better{2.90} & $\cmark$ \\ \cline{2-8}
    LP$_{ndet}^{loops}$ & $\forall\forall\Etau$ & 35 & 70   & 0.61 & TO & 4231.05 & \better{246.30} & $\cmark$ \\ \cline{2-8}
    LP$_{ndet}^{bugs}$ & $\forall\forall\Etau$ & 17 & 34   & 0.13 & TO & 30.86 & \better{16.99} & $\xmark$ \\ \cline{1-8}
    EFLP & $\forall\forall\Etau$ & 32 & 64 & 0.42 & \better{166.28} & {22.58} & \rustnum{TO} & $\cmark$ \\ \cline{2-8}
    EFLP$_{ndet}$ & $\forall\forall\Etau$ & 22 & 44   & 0.19 & TO & 160.30 & \better{40.74} & $\cmark$  \\ \cline{1-8}
    % EFLP_{ndet}^{loops} & 45 & 90 & 0.03 & 1.02 & TO & 221.47 & \rustnum{TO} \\ \cline{1-8}
    CacheTA & $\forall\forall\Atau\Etau$ & 13 & 26   & 0.21 &  TO & {2.40} & \better{1.18} & $\xmark$ \\ \cline{1-8}
    % CacheTA$_{ndet}$ & 58 & 116   & 4.81 & TO & 4.30 & \rustnum{TO} \\ \cline{1-8}
    % CacheTA$_{ndet}^{loops}$ & 35 & 70   & 1.86 & TO & 159.07 & \rustnum{TO} \\ \cline{1-8}
\end{tabular}

%% file: tables/table_loop.tex
\begin{tabular}{llcccc}
\toprule
\textbf{Type} & \textbf{Cases} & 
\textbf{Encode} & 
\textbf{Solving} & 
\textbf{\ourtool} & 
\textbf{\cite{hssb23}} \\
\midrule
\multirow{5}{*}{$\aesim$} 
 & ABP            & 0.27 & 0.01 & \better{0.28} & 9.37  \\
 & ABP w/bug     & 0.35 & 0.01  & \better{0.36} & 9.46 \\
 & MM             & 0.45 & 0.02  & \better{0.47}  & 67.74 \\
 & MM w/bug      & 0.45 & 0.01  & \better{0.46}  & 66.85 \\
 & CBF            & 0.20 & 0.01  & \better{0.21}  & 3.49 \\
 & CBF w/bug      & 0.19  & 0.01   & \better{0.20}  & 3.51 \\
\midrule
\multirow{4}{*}{$\easim$}
 & RP             & 0.20 & 0.32   & \better{0.52}   & 1.09 \\
 & RP no sol.   & 0.18 & 0.28   & \better{0.46}   & 1.02 \\
 & GCW            & 0.12   & 0.29   & \better{0.42} & 3.36 \\
 & GCW no sol.     & 0.11   & 0.27   & \better{0.39}     & 2.27 \\
\bottomrule
\end{tabular}

%% file: related.tex
\section{Related Work}
\label{sec:related}

There has been a recent surge of model checking techniques for
hyperproperties~\cite{cfst19,fht18,frs15,cfkmrs14}, employing various techniques (e.g., alternating automata, model counting, strategy synthesis, etc).
To our knowledge, there are only two existing model checkers for hyperproperties: \MCHyper~\cite{cfkmrs14} and \AutoHyper~\cite{bf23}.
The tool \MCHyper (which takes input models in \verilog) implements some of these ideas by computing the self-composition of the input model and reducing the problem to LTL model checking on top of the model checker \ABC~\cite{brayton10abc}.
However, these efforts generally fall short in proposing a general push-button method to deal with identifying bugs with respect to \HyperLTL formulas involving arbitrary quantifier alternation.
That is, \MCHyper requires external intervention to eleminate trace quantifier alternations, such as a game-based method. Thus, there is no basis of comparison between \HyperQB and \MCHyper.

As discussed throughout the paper, a more recent model checker \AutoHyper~\cite{bf23}, written in F\#, is an {\em explicit-state} tool and implements an automata-based verification approach based on language complementation and inclusion. 
It supports full \HyperLTL and is complete for properties with arbitrary quantifier alternations.
\AutoHyper is capable of generating counterexamples, but the performance may degrade significantly.
As shown in detail in~\Cref{sec:eval}, \HyperQB significantly outperforms \AutoHyper.
Having said that, \AutoHyper by design is better suited to carry out full verification, while \HyperQB is a BMC and inherently better to identify bugs quickly.
Finally, \AutoHyper cannot verify specifications in \AHLTL (i.e., it is incapable of handling all cases we presented in~\Cref{tab:result-ahltl}).

%The model checker \code{MCHyper}~\cite{frs15,cfst19,bcbfs21} processes Verilog designs and was initially developed to support alternation-free fragments of HyperLTL. 
%
%Its capabilities were later extended to handle a single quantifier alternation and asynchronous properties. 
%
%In contrast, \HyperQB provides greater expressive power by supporting arbitrary quantifier alternations and directly verifying asynchronous hyperproperties.
%
%Other verification methodologies are also employed. 
%
Finally, bounded model checkers, such as \code{EBMC}~\cite{KroeningEBMC}, accept both \verilog and \code{SystemVerilog} inputs for the verification of LTL properties.
Additionally, specialized tools exist to check for specific categories of information-flow properties. 
For instance, \code{IODINE}~\cite{Gleissenthall2019iodine} verifies constant-time execution on \verilog designs. 
Similarly, \code{SYLVIA}~\cite{Ryan2023Sylvia}, a symbolic execution engine, is used to identify security vulnerabilities, a subset of which includes observational determinism properties.

%% file: concl.tex
\section{Conclusion and Future Work}
\label{sec:concl}

We introduced the tool \ourtool, a QBF/SMT-based bounded model checker for \HyperLTL and \AHLTL, which allows input models in the \nusmv and \verilog languages.
\HyperQB implements four different semantics that ensure the
soundness of inferring the outcome of the model checking problem.
\HyperQB enjoys user interfaces for command-line access as well as GUIs for standalone and web-based applications.
The internal decision procedure is based on SMT and QBF solving.
Through a rich set of case studies, we demonstrated the effectiveness
and efficiency of \HyperQB in the verification of information-flow
properties, linearizability in concurrent data structures, path
planning in robotics, and fairness in non-repudiation protocols.

We are currently extending \HyperQB to accept \code{C} programs as input.
We are also developing an API  to provide the core functionality of the tool to other developers. 

%% file: appendix.tex
\section{Full Grammar for Input Formulas}\label{appendix:fullgrammar}

\Cref{fig:fullgrammar} presents the full grammar of \HyperLTL and \AHLTL formulas as implemented in our Rust-based parser. 
It enables \HyperQB to uniformly parse formulas from standard \HyperLTL to its asynchronous extension \AHLTL, supporting nested quantifiers and mixed temporal operators.
Implemented using the \texttt{pest} parsing library, the grammar remains modular, maintainable, and easily extensible to new logical constructs.

\begin{figure}
    \centering
\begin{lstlisting}
    WHITESPACE = _{ " " | "\t" | "\r\n" | "\n" }
    
    path_formula = { ("Forall" | "Exists") ~ ident ~ "." ~ form_rec }
    form_rec = { path_formula | traj_formula | inner_hltl }
    traj_formula = { ("A" | "E") ~ ident ~ "." ~ ahltl_form_rec }
    ahltl_form_rec = { traj_formula | inner_altl }
    
    inner_hltl = { hequal } 
    inner_altl = { aequal }
    
    hequal = { himpl ~ ("=" ~ hequal)? }
    himpl = { hdisj ~ ("->" ~ himpl)? }
    hdisj = { hconj ~ ("|" ~ hdisj)? }
    hconj = { huntl ~ ("&" ~ hconj)? }
    huntl = { hrels ~ ("U" ~ huntl)? }
    hrels = { hfactor ~ ("R" ~ hrels)? }
    hfactor = { unop ~ hfactor | "(" ~ inner_hltl ~ ")" | hltl_atom }
    hltl_atom = { ident ~ "[" ~ ident ~ "]" | constant | number }
    
    aequal = { aimpl ~ ("=" ~ aequal)? }
    aimpl = { adisj ~ ("->" ~ aimpl)? }
    adisj = { aconj ~ ("|" ~ adisj)? }
    aconj = { auntl ~ ("&" ~ aconj)? }
    auntl = { arels ~ ("U" ~ auntl)? }
    arels = { afactor ~ ("R" ~ arels)? }
    afactor = { aunop ~ afactor | "(" ~ inner_altl ~ ")" | altl_atom }
    altl_atom = { ident ~ "[" ~ident~ "]" ~ "[" ~ ident ~ "]" | constant | number}
    
    aunop = { "G" | "F" | "~" }
    unop = { "G" | "F" | "X" | "~" }
    ident = @{ ASCII_ALPHA ~ (ASCII_ALPHANUMERIC | "_" | ".")* }
    constant = {"TRUE" | "FALSE"}
    number = @{ ASCII_DIGIT+ | "#b" ~ ("0" | "1")+ }
    
    formula = _{ SOI ~ path_formula ~ EOI }
\end{lstlisting}
    \caption{Grammar of HyperLTL and A-HLTL input formulas.}
    \label{fig:fullgrammar}
\end{figure}

%% file: main.bbl
\newcommand{\SortNoop}[1]{}
\begin{thebibliography}{10}

\bibitem{mermaid}
\url{https://mermaid.js.org/}.

\bibitem{2005Verilog}
{IEEE} standard for verilog hardware description language.

\bibitem{bk08-book}
C.~Baier and J-.P. Katoen.
\newblock {\em Principles of Model Checking}.
\newblock The MIT Press, 2008.

\bibitem{bcbfs21}
J.~Baumeister, N.~Coenen, B.~Bonakdarpour, B.~Finkbeiner, and C.~S{\'{a}}nchez.
\newblock A temporal logic for asynchronous hyperproperties.
\newblock In {\em The 33rd International Conference on Computer Aided
  Verification (CAV)}, pages 694--717, 2021.

\bibitem{bf23}
R.~Beutner and B.~Finkbeiner.
\newblock {A}uto{H}yper: Explicit-state model checking for {H}yper{LTL}.
\newblock In {\em Proceedings of the 29th International Conference on Tools and
  Algorithms for the Construction and Analysis of Systems(TACAS)}, pages
  145--163, 2023.

\bibitem{bss18}
B.~Bonakdarpour, C.~S{\'{a}}nchez, and G.~Schneider.
\newblock Monitoring hyperproperties by combining static analysis and runtime
  verification.
\newblock In {\em Proceedings of the 8th Leveraging Applications of Formal
  Methods, Verification and Validation (ISoLA)}, pages 8--27, 2018.

\bibitem{brayton10abc}
R.~K. Brayton and A.~Mishchenko.
\newblock {ABC:} an academic industrial-strength verification tool.
\newblock volume 6174 of {\em LNCS}, pages 24--40. Springer, 2010.

\bibitem{cimatti1999nusmv}
Alessandro Cimatti, Edmund Clarke, Fausto Giunchiglia, and Marco Roveri.
\newblock Nusmv: A new symbolic model verifier.
\newblock In {\em International conference on computer aided verification},
  pages 495--499. Springer, 1999.

\bibitem{cbrz01}
E.~M. Clarke, A.~Biere, R.~Raimi, and Y.~Zhu.
\newblock Bounded model checking using satisfiability solving.
\newblock {\em Formal Methods in System Design}, 19(1):7--34, 2001.

\bibitem{cfkmrs14}
M.~R. Clarkson, B.~Finkbeiner, M.~Koleini, K.~K. Micinski, M.~N. Rabe, and
  C.~S{\'{a}}nchez.
\newblock Temporal logics for hyperproperties.
\newblock In {\em Proceedings of the 3rd Conference on Principles of Security
  and Trust {POST}}, pages 265--284, 2014.

\bibitem{cs10}
M.~R. Clarkson and F.~B. Schneider.
\newblock Hyperproperties.
\newblock {\em Journal of Computer Security}, 18(6):1157--1210, 2010.

\bibitem{cfst19}
N.~Coenen, B.~Finkbeiner, C.~S{\'{a}}nchez, and L.~Tentrup.
\newblock Verifying hyperliveness.
\newblock In {\em Proceedings of the 31st International Conference on Computer
  Aided Verification (CAV)}, pages 121--139, 2019.

\bibitem{dmb08}
L.~M. de~Moura and N.~Bj{\o}rner.
\newblock Z3: An efficient {SMT} solver.
\newblock In {\em Tools and Algorithms for the Construction and Analysis of
  Systems (TACAS)}, pages 337--340, 2008.

\bibitem{fht18}
B.~Finkbeiner, C.~Hahn, and H.~Torfah.
\newblock Model checking quantitative hyperproperties.
\newblock In {\em Proceedings of the 30th International Conference on Computer
  Aided Verification}, pages 144--163, 2018.

\bibitem{frs15}
B.~Finkbeiner, M.~N. Rabe, and C.~S{\'{a}}nchez.
\newblock Algorithms for model checking {H}yper{LTL} and {H}yper{CTL}*.
\newblock In {\em Proceedings of the 27th International Conference on Computer
  Aided Verification (CAV)}, pages 30--48, 2015.

\bibitem{Gleissenthall2019iodine}
Klaus~V. Gleissenthall, Rami~G\"{o}khan Kici, Deian Stefan, and Ranjit Jhala.
\newblock Iodine: verifying constant-time execution of hardware.
\newblock In {\em Proceedings of the 28th USENIX Conference on Security
  Symposium}, SEC'19, page 1411–1428. USENIX Association, 2019.

\bibitem{YosysHQ}
YosysHQ GmbH.
\newblock Yosys open synthesis suite.

\bibitem{gm82}
J.~A. Goguen and J.~Meseguer.
\newblock Security policies and security models.
\newblock In {\em IEEE Symp. on Security and Privacy}, pages 11--20, 1982.

\bibitem{hw90}
M.~Herlihy and J.~M. Wing.
\newblock Linearizability: {A} correctness condition for concurrent objects.
\newblock {\em {ACM} Transactions on Programming Languages and Systems},
  12(3):463--492, 1990.

\bibitem{hbfs23}
T.-H. Hsu, B.~Bonakdarpour, B.~Finkbeiner, and C.~S{\'{a}}nchez.
\newblock Bounded model checking for asynchronous hyperproperties.
\newblock In {\em Proceedings of the 29th International Conference on Tools and
  Algorithms for Construction and Analysis of Systems (TACAS)}, pages 29--46,
  2023.

\bibitem{hssb23}
T.-H. Hsu, C.~S{\'{a}}nchez, , S.~Sheinvald, and B.~Bonakdarpour.
\newblock Efficient loop conditions for bounded model checking hyperproperties.
\newblock In {\em Proceedings of the 29th International Conference on Tools and
  Algorithms for Construction and Analysis of Systems (TACAS)}, pages 66 -- 84,
  2023.

\bibitem{hsb21}
T.-H. Hsu, C.~S{\'{a}}nchez, and B.~Bonakdarpour.
\newblock Bounded model checking for hyperproperties.
\newblock In {\em Proceedings of the 27th International Conference on Tools and
  Algorithms for the Construction and Analysis of Systems (TACAS)}, pages
  94--112, 2021.

\bibitem{KroeningEBMC}
Daniel Kroening.
\newblock Ebmc.

\bibitem{Ryan2023Sylvia}
Kaki Ryan and Cynthia Sturton.
\newblock Sylvia: Countering the path explosion problem in the symbolic
  execution of hardware designs.
\newblock In {\em Proceedings of the 23rd Conference on Formal Methods in
  Computer-Aided Design (FMCAD)}, FMCAD, pages 110--121, 2023.

\bibitem{t19}
L.~Tentrup.
\newblock {CAQE} and quabs: Abstraction based {QBF} solvers.
\newblock {\em Journal of Satisfiability Boolean Modeling and Computation},
  11(1):155--210, 2019.

\bibitem{wzbp19}
Y.~Wang, M.~Zarei, B.~Bonakdarpour, and M.~Pajic.
\newblock Statistical verification of hyperproperties for cyber-physical
  systems.
\newblock {\em {ACM} Transactions on Embedded Computing Systems},
  18(5s):92:1--92:23, 2019.

\end{thebibliography}
